\documentclass{article}
\pdfoutput=1
\usepackage{epsfig}
\usepackage{inputenc}
\usepackage{graphicx}
\usepackage{color}
\usepackage{array}
\usepackage{pstricks}
\usepackage{titlesec}
\usepackage{makeidx}
\usepackage{amsmath,amssymb,mathrsfs}
\usepackage{bm}
\usepackage{pict2e}
\usepackage{supertabular}
\usepackage{natbib}
\usepackage[a4paper,pdftex,verbose,margin=2.5cm,twoside]{geometry}

\begin{document}


\title{On the effect of rotation on magnetohydrodynamic turbulence at high magnetic Reynolds number}

\author{B.F.N. Favier${\dag}$$^{\ast}$\thanks{$^\ast$Corresponding author. Email: benjamin.favier@ncl.ac.uk
\vspace{6pt}}, F.S. Godeferd${\dag}$ and C. Cambon${\dag}$ \\ \\ ${\dag}$Universit\'e de Lyon, Laboratoire de M\'ecanique des Fluides et d'Acoustique, UMR 5509,\\
\'Ecole Centrale de Lyon, CNRS, UCBL, INSA F-69134 Ecully Cedex, France 
}

\maketitle

\begin{abstract}
This article is focused on the dynamics of a rotating electrically conducting fluid in a turbulent state.
As inside the Earth's core or in various industrial processes, a flow is altered by the presence of both background rotation and a large scale magnetic field.
In this context, we present a set of three-dimensional direct numerical simulations of incompressible decaying turbulence.
We focus on parameters similar to the ones encountered in geophysical and astrophysical flows, so that the Rossby number is small, the interaction parameter is large, but the Elsasser number, defining the ratio between Coriolis and Lorentz forces, is about unity.
These simulations allow to quantify the effect of rotation and thus inertial waves on the growth of magnetic fluctuations due to Alfv\'en waves.
Rotation prevents the occurence of equipartition between
kinetic and magnetic energies, with a reduction of magnetic
energy at decreasing Elsasser number $\Lambda$.
It also causes a decrease of energy transfer mediated
by cubic correlations.
In terms of flow structure, a decrease of $\Lambda$
corresponds to an increase in the misalignment of velocity
and magnetic field.\bigskip

\end{abstract}

\section{Introduction}
Turbulence in electrically conducting fluids plays a role in many industrial processes and in flows from geophysical to astrophysical interests.
In addition to containing a wide distribution of spatial and temporal scales, these flows are often strongly anisotropic due to the effect of solid-body rotation, of a buoyancy gradient or of an imposed magnetic field. Independently, these forces and their effects on the anisotropy of turbulence have received an increasing interest.

The case of rotating turbulence of non-conducting fluid has been extensively studied \citep{jacq90,CAMBON-MANSOUR-GODEFERD,mori01}.
The understanding of the dynamics and anisotropy of homogeneous rotating turbulence has led to many advances in atmospheric and geophysical flows.
A quick description of the main results can be drawn as follows (for more details, see Sagaut and Cambon 2008): the energy cascade is inhibited by the Coriolis force, so that the dissipation rate is reduced \citep{bard85,jacq90}; a transition from 3D to quasi-2D state develops and coherent vortical structures elongated in the direction of the rotation axis break the cyclone-anticyclone symmetry.

A large scale magnetic field exists in many astrophysical objects (such as star and planet interiors or solar wind) or industrial configurations.
The temporal and spatial variations of this imposed field may be neglected so that, at least locally, one may assume that it is uniform and steady \citep{moff71}.
The transformation of initially isotropic turbulence by a static magnetic field is also a well documented problem, from theoretical studies \citep{lehn55,moff67} to numerical \citep{schu76,ough94,zika98,knae04,voro05,bigo08,teac09} and experimental works \citep{alem79}.
The key phenomena are: a strong reduction of nonlinear transfers of energy along the direction of the imposed magnetic field, eventually leading to a quasi-twodimensional state and depending on the value of the imposed magnetic field; and anisotropic Ohmic dissipation whose relative amplitude depends on the magnetic Reynolds number.

In realistic flows, more than one force is usually responsible for anisotropic phenomena.
The case of rotating stratified turbulence has been studied because of its direct application to atmospheric and oceanic flows \citep{camb01,liec05}.
However, it seems that the coupled effect of the Coriolis force and the Lorentz force on homogeneous turbulence has received too few interest, except for theoretical considerations \citep{lehn54,lehn55,moff70}.

In this paper, we present the results of numerical simulations of incompressible homogeneous magnetohydrodynamic (MHD) turbulence submitted to a uniform magnetic field and to solid-body rotation.
In order to avoid introducing additional parameters to the already complex parameter space (Reynolds numbers, Rossby number, Elsasser number, interaction parameter, spectral cut-off, etc), we consider the freely decaying turbulent regime, that is without introducing forcing terms.
We also consider the simplest configuration, in which the imposed magnetic field is aligned with the rotation axis so that the configuration is axisymmetric.
We also select dimensionless numbers that correspond to regimes comparable to the ones observed in many geophysical and astrophysical systems: the Rossby number is very low (\textit{i.e.} the rotation has a strong effect on the dynamics), the interaction parameter is moderate (\textit{i.e.} the Lorentz force also has an important effect, but the regime is however far from wave turbulence), and the Elsasser number, characterizing the relative strength of the Coriolis and Lorentz forces, is about unity.
Note however that the Lehnert number (see below) is always small compare to unity so that we consider flows dominated by rotation and inertial waves.

In the following, we present the equations and nondimensional parameters in section 2, then the numerical resolution method in section 3, used to obtain the linear dynamics results presented in section 4. Nonlinear simulations are discussed in section 5 starting with the characterization or rotating MHD turbulence in physical space, followed by an extensive discussion of the dynamics in spectral space. Conclusions are drawn in section 6.
%
%
\section{Governing equations and dimensionless numbers}
\label{sec:eq}
We consider the homogeneous turbulent flow of an incompressible electrically conducting fluid. The fluid is characterized by its kinematic viscosity $\nu$ and its magnetic diffusivity $\eta$; with $\eta=(\sigma\mu_0)^{-1}$, $\sigma$ the electrical conductivity and $\mu_0$ the magnetic permeability.
The Navier-Stokes equations for an incompressible fluid, including the Coriolis and Lorentz forces, are written in the rotating frame as:
\begin{align}
\label{eq:momentum}
\frac{\partial \bm{u}}{\partial t}-\nu\nabla^2\bm{u} & =\bm{u}\times\left(\bm{\omega}+2\bm{\Omega}\right)+\bm{j}\times\left(\bm{B}_0+\bm{b}\right)-{\boldsymbol{\nabla}} P\\
\label{eq:div}
{\boldsymbol{\nabla}}{\boldsymbol{\cdot}}\bm{u} & =0
\end{align}
where $\bm{\omega}={\boldsymbol{\nabla}}\times\bm{u}$ is the vorticity, $\bm{b}$ is the fluctuating magnetic field, $\bm{j}={\boldsymbol{\nabla}}\times\bm{b}$ is the normalized electrical current and $P$ is the pressure modified by magnetic pressure and centrifugal terms.
Within the magnetohydrodynamic theory, the fluctuating magnetic field is derived from the induction equation
\begin{equation}
\label{eq:induction}
\frac{\partial \bm{b}}{\partial t}-\eta\nabla^2\bm{b}={\boldsymbol{\nabla}}\times\left(\bm{u}\times\left(\bm{B}_0+\bm{b}\right)\right)
\end{equation}
with the additional constraint
\begin{equation}
\label{eq:divb}
{\boldsymbol{\nabla}}{\boldsymbol{\cdot}}\bm{b}=0 \ .
\end{equation}

The initial state is fully developed and isotropic turbulence without magnetic fluctuations. The initial root mean square velocity is $u_0$ and the initial integral length scale is $l_0$. The Reynolds number $Re$ and its magnetic counterpart $R_M$ are defined by
\begin{equation}
Re=\frac{u_0l_0}{\nu} \quad \textrm{and} \quad R_M=\frac{u_0l_0}{\eta} \ .
\end{equation}
We focus on developed turbulent velocity field at large Reynolds number.
The magnetic Reynolds number is also assumed large, with a magnetic Prandtl number of about one for all our simulations.
Cases ranging from moderate to low magnetic Reynolds number, of specific interest for industrial applications, will be considered in future studies.

The flow is submitted to a uniform magnetic field $\bm{B}_0$ (all the magnetic quantities are given in Alfv\'en-speed unit such that $\bm{B}_0=\bm{B}/\sqrt{\rho_0\mu_0}$ where $\rho_0$ is the density of the fluid) in a frame rotating with a constant angular velocity $\bm{\Omega}$.
Both $\bm{B}_0$ and $\bm{\Omega}$ are vertical.
The present axisymmetric case is the simplest configuration since all statistical quantities depend only on their orientations with respect to the direction of symmetry (the vertical one here).
In the Earth's core, the strong differential rotation between the inner core and the mantle causes the magnetic field to be mainly toroidal, thus locally perpendicular to the rotation axis.
However, this perpendicular case is much more complex in terms of anisotropy and is not considered here.
The ratio between the eddy turnover time $l_0/u_0$ and the magnetic damping time $\eta/B_0^2$ is the magnetic interaction number
\begin{equation}
N=\frac{B_0^2l_0}{\eta u_0} \ .
\end{equation}
This parameter can also be seen as the ratio between the amplitude of the Lorentz force with respect to the advective term in equation \eqref{eq:momentum}.
In this paper, the intensity of the imposed magnetic field remains constant.
We focus on the dynamics at moderate interaction paramter ($N=12$, see below).
Note however that we still consider the so-called strong turbulence limit where $B_0\approx u_0$, so that the anisotropy resulting from the applied magnetic field is small.
The weak turbulence limit $B_0\gg u_0$ has been considered in the non-rotating case \citep{bigo08,alex07}, and one observes an inhibition of nonlinear kinetic energy transfers in the direction of the imposed magnetic field.
The effect of the rotation on this weak turbulence state will be considered in future studies.

The ratio between the eddy turnover time and the Coriolis parameter is the Rossby number
\begin{equation}
Ro=\frac{u_0}{2\Omega l_0} \ .
\end{equation}
In order to assess the effect of the Coriolis force on strong MHD turbulence, we consider different values of the rotation rate but we focus on the dynamics dominated by rotation so that the Rossby number is always small compared to unity.
The product between the above two dimensionless parameters defines the Elsasser number
\begin{equation}
\Lambda=\frac{B_0^2}{2\Omega\eta}\ ,
\end{equation}
characterizing the respective influence of the Coriolis force and the Lorentz force. For non-rotating MHD turbulence, $\Lambda\rightarrow\infty$ and we consider in the following Elsasser numbers down to $0.5$. Note that it is believed that the Elsasser number of the Earth's core is about unity.
The relative importance of the rotation and of the magnetic field can also be quantified by the Lehnert number \citep{lehn55}, which is the ratio between the frequency of Alfv\'en waves and the frequency of inertial waves:
\begin{equation}
\mathcal{L}=\frac{B_0}{2\Omega l_0} \ .
\end{equation}
In the present simulations, the Lehnert number is always small compared to one so that inertial waves are rapid compared to Alfv\'en waves. To reach larger values of the Elsasser number, one has to increase the value of the imposed magnetic field, thus considering the weak turbulence limit.
We also introduce a scale-dependent Lehnert number, based on a wave number $k$, as
\begin{equation}
\label{eq:lehnert}
\hat{\mathcal{L}}=\frac{B_0k}{2\Omega} \ .
\end{equation}
\begin{table}
\begin{center}
\caption{Set of parameters of the initial condition for anisotropic DNS. $R_{\lambda}$ is the Reynolds number based on the Taylor microscale.}
\begin{tabular}{cccccc}
\hline
   Resolution
  &  $u_0$
         & $l_0$
         & $\nu$ & $R_{\lambda}$ & $k_{\textrm{max}}l_{\eta}$ \\
\hline
   $256^3$ & 0.78 & 0.62 & 0.0025
         & 72 & 2.35 \\
\hline
  \end{tabular}
\label{tab:initDNS}
\end{center}
\end{table}
%
%
\section{Numerical method}
\label{sec_numerics}
A Fourier pseudo-spectral method is used to solve equations \eqref{eq:momentum} to \eqref{eq:divb}.
The velocity field and the fluctuating magnetic field are computed in a cubic box of side $2\pi$ with periodic boundary conditions and a resolution of $256^3$ Fourier modes.
A spherical truncation of Fourier components is used to remove completely the aliasing and the time scheme is third-order Adams-Bashforth.
The dissipative terms (proportional to $\nu$ and $\eta$ in \eqref{eq:momentum} and \eqref{eq:induction}) are implicitly solved through the change of variables
$\hat{\bm{u}}(\bm{k},t)\leftrightarrow \hat{\bm{u}}(\bm{k},t)\exp\left(-\nu k^2 t\right)$ and $\hat{\bm{b}}(\bm{k},t)\leftrightarrow \hat{\bm{b}}(\bm{k},t)\exp\left(-\eta k^2 t\right)$
where $\hat{\bm{u}}(\bm{k},t)$ and $\hat{\bm{b}}(\bm{k},t)$ are the Fourier transforms of $\bm{u}(\bm{x},t)$ and $\bm{b}(\bm{x},t)$ respectively.
The nonlinear terms are computed in physical space, and then obtained in Fourier space by using fast Fourier transforms.

The initial Eulerian velocity field comes from an isotropic simulation in the purely hydrodynamic case with zero angular velocity, whose initial velocity field is a random superposition of Fourier modes with an energy distribution given by the narrow-band spectrum $E(k,t=0)\approx k^4\exp(-2(k/k_i)^2)$ \citep{ROGALLO:81}.
During the initial stage only, a large-scale forcing is applied until a statistical steady state of classical isotropic turbulence is reached. This forcing is applied for wavenumbers such that $2\leq k\leq4$.
The resolution is such that the minimum value for $k_{\textrm{max}}l_{\eta}$ is $2.35$ for all computations, where $l_{\eta}$ is the Kolmogorov length scale \citep{jime93}.
At the beginning of the anisotropic simulations, the magnetic field $\bm{B}_0$ and the angular velocity are suddenly switched on, the forcing is turned off, and there are no magnetic fluctuations. We choose to consider the case without initial magnetic fluctuations to reduce the already wide parametric space and to focus on the growth of magnetic fluctuations. If the flow contains initial magnetic fluctuations, one has to define additional parameters, such as the initial ratio between magnetic and kinetic energies or velocity/magnetic field cross-correlations.
Some characteristics of the initial condition used for anisotropic simulations are gathered in Table \ref{tab:initDNS}.
%
%
\section{Linear inviscid dynamics}
\label{sec:linear}
\begin{figure}
\unitlength 0.6mm
\begin{picture}(200,110)
        \put(100,-20){\includegraphics[height=120\unitlength]{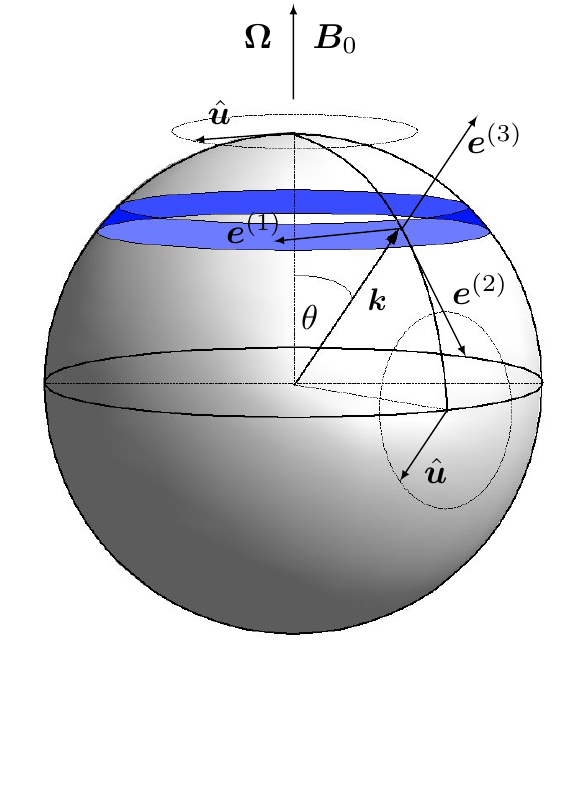}}
\end{picture}
\caption{(Color online) Craya-Herring frame $(\bm{e}^{(1)},\bm{e}^{(2)},\bm{e}^{(3)})$ in Fourier space.
An incompressible Fourier variable is perpendicular to the local wave vector $\bm{k}$ so that its contribution along $\bm{e}^{(3)}$ is zero.
Fourier modes in the shaded region contribute to  $E(k,\theta)$ (see equation \eqref{eq:ekt_def2}). The polar modes ($\theta\simeq 0$) contribute to horizontal kinetic energy, whereas equatorial modes ($\theta\simeq\pi/2$) contribute to both vertical (along $\bm{e}^{(2)}$) and horizontal (along $\bm{e}^{(1)}$) energies.
}%
\label{fig:spang_lukas}
\end{figure}
Let us first consider the inviscid linear dynamics of the problem.
Some details about the following oscillating solutions can be found in \cite{lehn54} and \cite{moff71}.
It is convenient to introduce the so-called Craya-Herring frame (see figure \ref{fig:spang_lukas} and \cite{saga08} for the link to the toroidal-poloidal decomposition in physical space) defined by
\begin{equation}
\bm{e}^{(1)}=\frac{\bm{k}\times\bm{n}}{|\bm{k}\times\bm{n}|}, \quad \bm{e}^{(2)}=\frac{\bm{k}}{k}\times\bm{e}^{(1)} \quad \textrm{and} \quad \bm{e}^{(3)}=\frac{\bm{k}}{k} \ .
\end{equation}
Given that both velocity and magnetic fields are divergence-free, the components along the wave vector are zero so that $\hat{u}^{(3)}(\bm{k})=\hat{b}^{(3)}(\bm{k})=0$.
Neglecting nonlinear terms and dissipative effects in equations \eqref{eq:momentum} and \eqref{eq:induction}, one can write the following linear system in the Craya-Herring frame:
\begin{equation}
\frac{\partial}{\partial t} \begin{pmatrix} \hat{u}^{(1)} \\ \hat{u}^{(2)} \\ \hat{b}^{(1)} \\ \hat{b}^{(2)} \end{pmatrix} + \begin{pmatrix} 0 & -\omega_i & -\textrm{i}\omega_a & 0 \\ \omega_i & 0 & 0 & -\textrm{i}\omega_a \\ -\textrm{i}\omega_a & 0 & 0 & 0 \\ 0 & -\textrm{i}\omega_a & 0 & 0 \end{pmatrix} \begin{pmatrix} \hat{u}^{(1)} \\ \hat{u}^{(2)} \\ \hat{b}^{(1)} \\ \hat{b}^{(2)} \end{pmatrix}=0 \ ,
\label{eq:NSmhdrotspectral}
\end{equation}
where $\omega_i=2\Omega\cos\theta$ is the dispersion relation of inertial waves, $\omega_a=B_0k\cos\theta$ is the dispersion relation of Alfv\'en waves and $\theta$ is the polar angle between $\bm{\Omega}$ (or $\bm{B}_0$) and the wave vector $\bm{k}$.
The dispersion relation of magneto-inertial waves follows from the eigenvalue problem, and write
\begin{equation}
\omega=\frac{1}{2}\left(\pm\omega_i\pm\sqrt{\omega_i^2+4\omega_a^2}\right) \ .
\label{eq:eigenvalues_br}
\end{equation}
\begin{figure}
\unitlength 1mm
\begin{picture}(200,100)
        \put(10,-7){\includegraphics[height=100\unitlength]{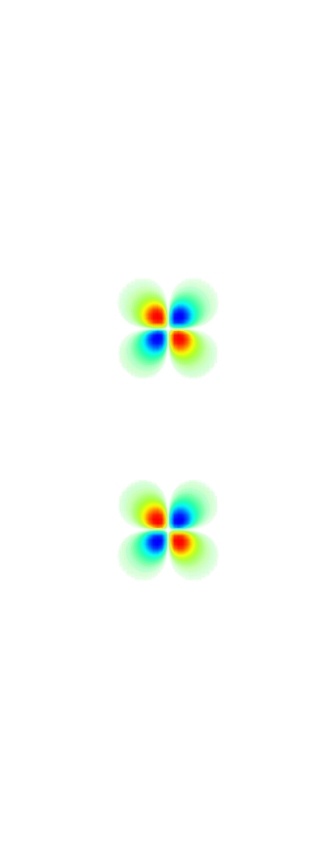}}
        \put(45,-7){\includegraphics[height=100\unitlength]{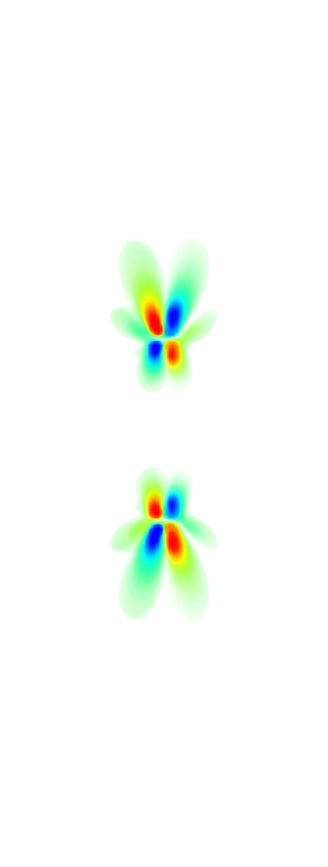}}
        \put(85,-7){\includegraphics[height=100\unitlength]{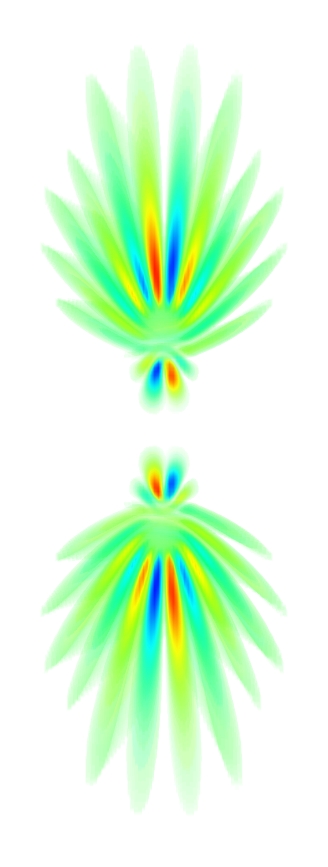}}
        \put(130,-7){\includegraphics[height=100\unitlength]{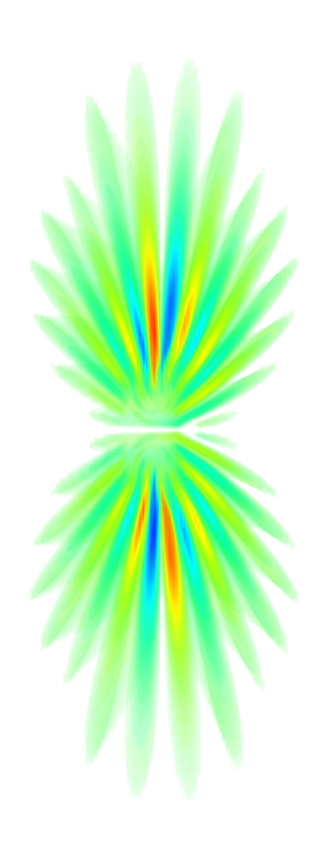}}
%
        \put(29.5,63){\vector(0,1){10}}
	\put(32,68){$\bm{B}_0$}
        \put(64.5,63){\vector(0,1){10}}
	\put(68,68){$\bm{B}_0$}
	\put(57,68){$\bm{\Omega}$}
        \put(148.8,63){\vector(0,1){10}}
	\put(141.3,68){$\bm{\Omega}$}
        \put(20,15){(a)}
	\put(55,15){(b)}
	\put(88,5){(c)}
	\put(133,5){(d)}
        \put(147.1,42.4){\large{$\bullet$}}
        \put(102.8,42.4){\large{$\bullet$}}
        \put(62.8,42.4){\large{$\bullet$}}
        \put(28.3,42.4){\large{$\bullet$}}
        \multiput(18,43.5)(4.,0){36}
        {\line(1,0){2}}
        \put(0,50){\includegraphics[height=30\unitlength]{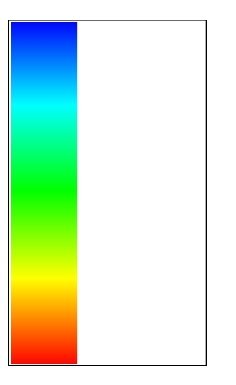}}
        \put(7,53.5){\scriptsize{$-u_z^{\textrm{max}}$}}
        \put(8,75){\scriptsize{$u_z^{\textrm{max}}$}}
\end{picture}
\caption[]{Visualizations of magneto-inertial wave packets. The vertical component $u_z$ of the velocity is presented in a vertical plan containing the initial impulse. $u_z^{\textrm{max}}$ is the maximum value of the vertical velocity. The dots indicate the vertical position of the initial perturbation. All visualizations are presented at $tB_0/l_f\approx6$ where $l_f$ is the characteristic length scale of the initial perturbation. (a) Pure Alfv\'en waves ($B_0=1$ and $\Omega=0$). (b) Magneto-inertial waves with $B_0=1$ and $\Omega=3$. (c) Magneto-inertial waves with $B_0=1$ and $\Omega=10$. (d) Pure inertial waves ($B_0=0$ and $\Omega=10$).}
\label{fig:visu_magneto_in}
\end{figure}
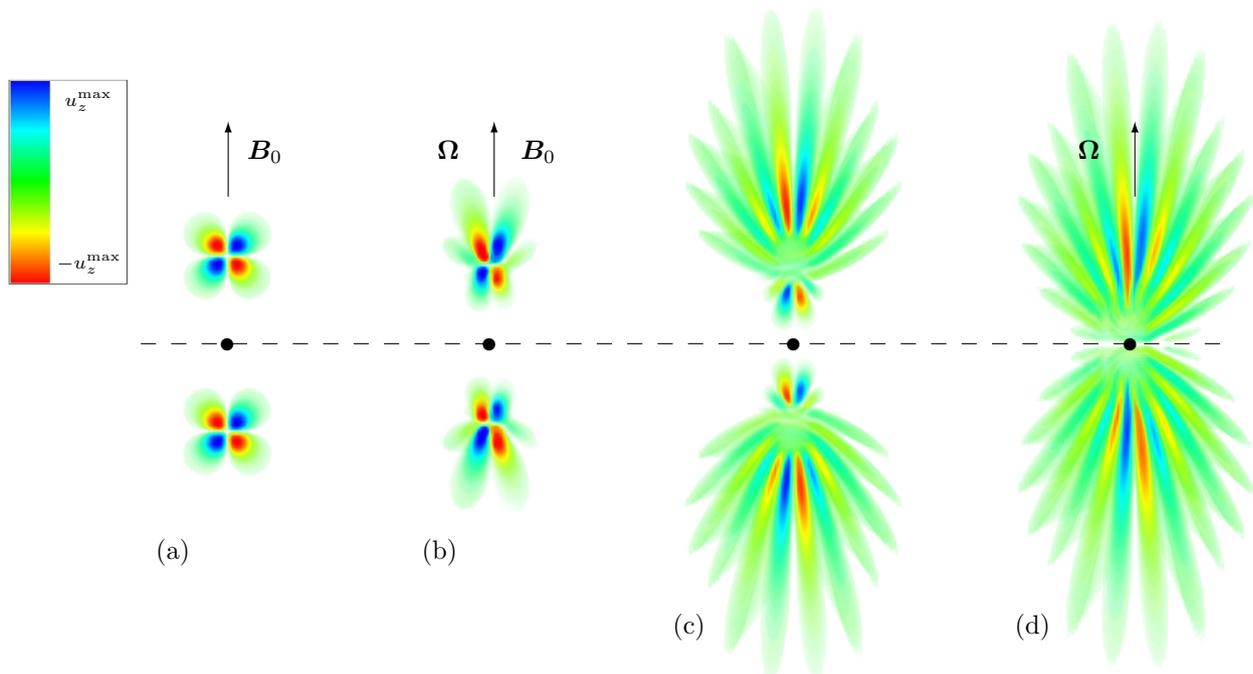
In the following, we assume that the Lehnert number is small so that $\omega_i\gg\omega_a$, \textit{i.e.} the characteristic frequency of inertial waves is much larger than the one of Alfv\'en waves. Note that this assumption is valid only for a certain range of scale since the Lehnert number defined by equation \eqref{eq:lehnert} depends on the wave number $k$. Using this assumption, and considering that the initial condition is $\hat{\bm{u}}(\bm{k},t=0)=\left(\hat{u}^{(1)}(\bm{k},0),\hat{u}^{(2)}(\bm{k},0)\right)$ for the velocity and $\hat{\bm{b}}(\bm{k},t=0)=(0,0)$ for the fluctuating magnetic field, the solutions are
\begin{align}
\label{eq:NSrotspectralsol1}
\begin{pmatrix} \hat{u}^{(1)}(\bm{k},t) \\ \hat{u}^{(2)}(\bm{k},t) \end{pmatrix} & =\begin{pmatrix} \cos\omega_it & -\sin\omega_it \\ \sin\omega_it & \cos\omega_it \end{pmatrix} \begin{pmatrix} \hat{u}^{(1)}(\bm{k},0) \\ \hat{u}^{(2)}(\bm{k},0) \end{pmatrix}\\
\begin{pmatrix} \hat{b}^{(1)}(\bm{k},t) \\ \hat{b}^{(2)}(\bm{k},t) \end{pmatrix} & =\mathrm{i}\frac{2\omega_a}{\omega_i}\sin(\omega_it/2)\begin{pmatrix} \cos(\omega_it/2) & \sin(\omega_it/2) \\ -\sin(\omega_it/2) & \cos(\omega_it/2) \end{pmatrix} \begin{pmatrix} \hat{u}^{(1)}(\bm{k},0) \\ \hat{u}^{(2)}(\bm{k},0) \end{pmatrix} \ .
\label{eq:NSrotspectralsol2}
\end{align}
\begin{figure}
\unitlength 0.4mm
\begin{picture}(250,140)
        \put(12,0){\includegraphics[height=140\unitlength]{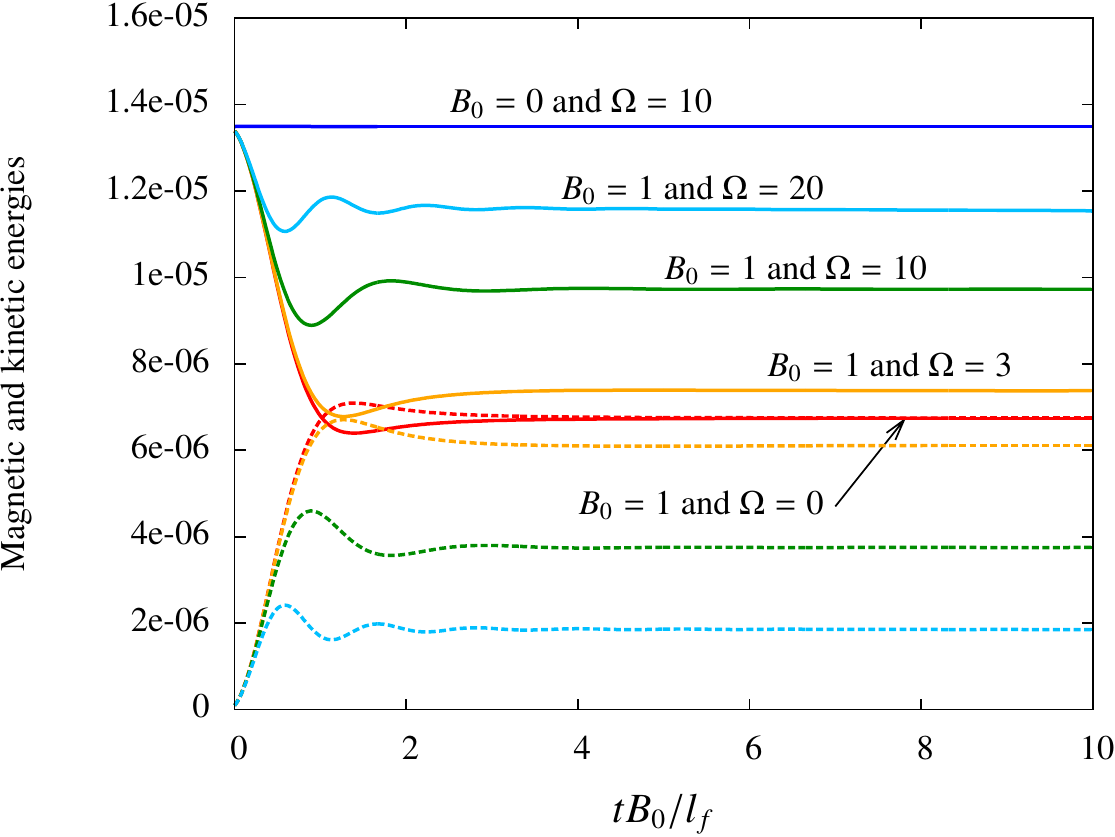}}
        \put(225,0){\includegraphics[height=140\unitlength]{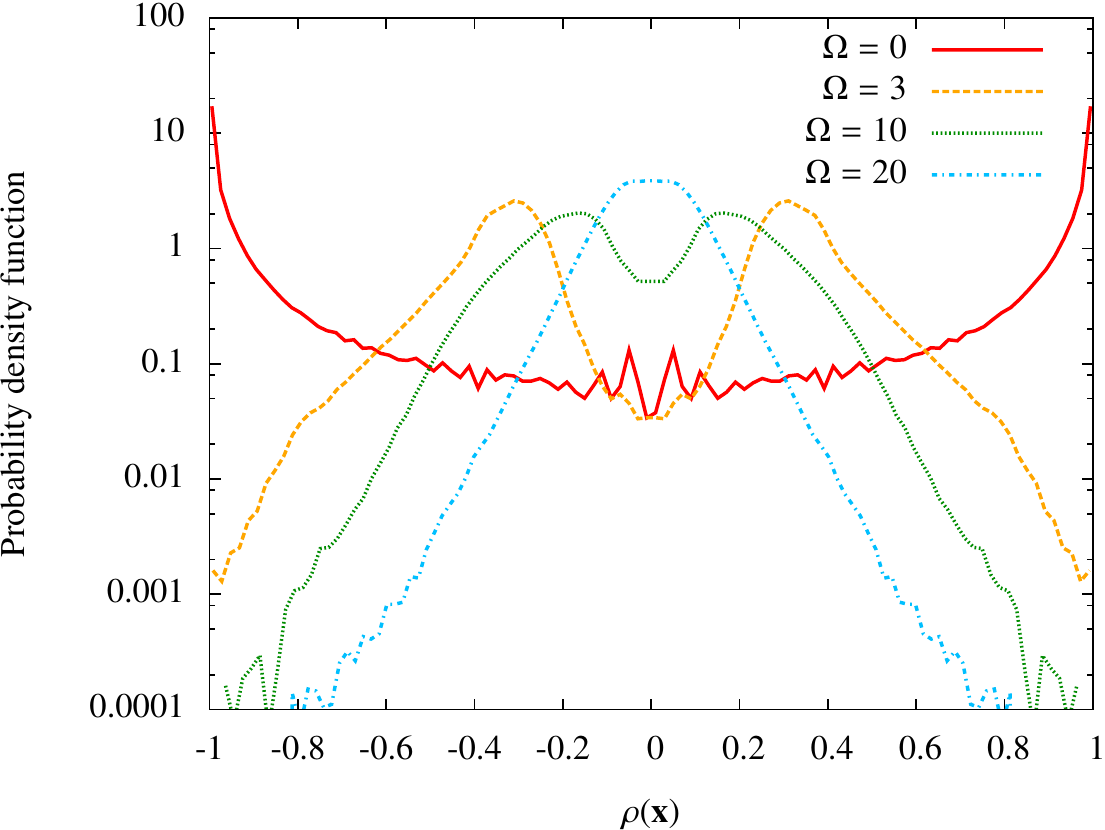}}
	\put(174,125){(a)}
	\put(260,125){(b)}
\end{picture}
\caption[]{(Color online) (a) Kinetic energy (continuous lines) and magnetic energy (dotted lines) for magneto-inertial waves. The time is normalized by $l_f/B_0$ where $l_f$ is the characteristic lenght scale of the initial impulse and $B_0$ is the imposed magnetic field (in Alfv\'en speed units).(b) Probability density function of the cross correlation \eqref{eq:corrub} at $tB_0/l_f\approx10$.}
\label{fig:linear_results}
\end{figure}
The time dependence of the velocity Fourier mode is characteristic of pure inertial waves: the Fourier mode rotates in the plane $(\bm{e}^{(1)},\bm{e}^{(2)})$ with an angular velocity fixed by the dispersion relation of inertial waves.
Concerning the magnetic mode, we first note that its amplitude is modulated by $\omega_a/\omega_i$, which is assumed to be very small here, so the magnetic energy should be very small compared to the kinetic energy when the rotation is dominant.
Secondly, it is known that, without rotation, the magnetic mode and the velocity mode are colinear (see for example Moffatt 1967) so that pure Alfv\'en waves are characterized by $\hat{\bm{u}}(\bm{k},t)\parallel\hat{\bm{b}}(\bm{k},t)$ (independently of the wave vector considered, so that this preferential alignment still holds in physical space).
With rotation, a phase shift appears between the magnetic and kinetic modes in spectral space, so that there is no reason to observe velocity fluctuations parallel to magnetic ones in physical space.

In the following, we propose a simple numerical experiment to assess qualitatively the previous linear statements.
Using the numerical method described above, we perform direct simulations of magneto-inertial waves, with a resolution of $128\times128\times256$ Fourier modes.
This resolution is well above what is needed for such simulation but it ensures that no spurious effects of the periodic boundary conditions appear.
From a non-magnetized fluid initially at rest, we introduce a small horizontal velocity perturbation (both $\bm{\Omega}$ and $\bm{B}_0$ are vertical) localized in space and in time to approximate the impulse response of the system.
The amplitude of the perturbation is very small so that nonlinear interactions are negligible throughout the simulation (nonlinear terms are nevertheless removed to ensure that the results are purely linear).
We set $B_0=1$ and we consider four rotation rates: $\Omega=0$ (pure Alfv\'en waves), $3$, $10$ and $20$, and we also perform a purely hydrodynamic simulation with $B_0=0$ and $\Omega=10$.
A volume rendering of the vertical component of the velocity is presented on figure \ref{fig:visu_magneto_in}. Without rotation, an upward and a downward Alfv\'en wave packets propagate along the imposed vertical magnetic field at a constant velocity. The shape of each wave packets is directly linked to the initial perturbation, and does not vary in time (dissipative effects are neglected).
When rotation is introduced, dispersion of the energy along non-vertical directions is observed, which is very similar to the one observed in purely inertial waves (see figure \ref{fig:visu_magneto_in}(d)).
It is well-known that the frequency of inertial waves depends only the angle $\theta$ between $\bm{\Omega}$ and $\bm{k}$.
In the case of a sinusoidal forcing \citep{mcewan70}, the ratio between the forcing frequency and the rotation rate sets the direction of propagation of energy.
In our case, all the available frequencies are excited by the initial pulse so that one observes different directions of propagation.
Figure \ref{fig:linear_results}(a) presents the evolution with time of both kinetic and magnetic energies. Without rotation, the equipartition between kinetic and magnetic energies is observed after some relaxation. As predicted by the linear solution \eqref{eq:NSrotspectralsol2}, the magnetic energy is strongly damped in dominant rotation cases. Finally, the probability density function of the cross-correlation between $\bm{u}(\bm{x})$ and $\bm{b}(\bm{x})$, defined as
\begin{equation}
\label{eq:corrub}
\rho(\bm{x})=\frac{2\left(\bm{u}(\bm{x}){\boldsymbol{\cdot}}\bm{b}(\bm{x})\right)}{\bm{u}^2(\bm{x})+\bm{b}^2(\bm{x})}
\end{equation}
is plotted in figure \ref{fig:linear_results}(b).
For pure Alfv\'en waves, the colinearity between $\bm{u}$ and $\bm{b}$ is clear and the rotation tends to break the preferential alignment.
For low value of the Lehnert number (\textit{i.e.} dominant rotation), it seems that the perpendicular state becomes more probable.

The previous linear observations are to be kept in mind for the following nonlinear study.
However, in order to be clear, we insist on the fact that the following MHD direct simulations are in the strong turbulence regime, meaning that nonlinearities are dominant with respect to the imposed magnetic field.
In that case, the preferential alignment between $\bm{u}$ and $\bm{b}$ will not be observed but the linear effect of the rotation could give some insights on the nonlinear results.
%
%
\section{Nonlinear simulation results}
\label{sec:induc}
\begin{figure}[b!]
\unitlength 0.6mm
\begin{picture}(200,120)
        \put(36,-4){\includegraphics[height=118\unitlength]{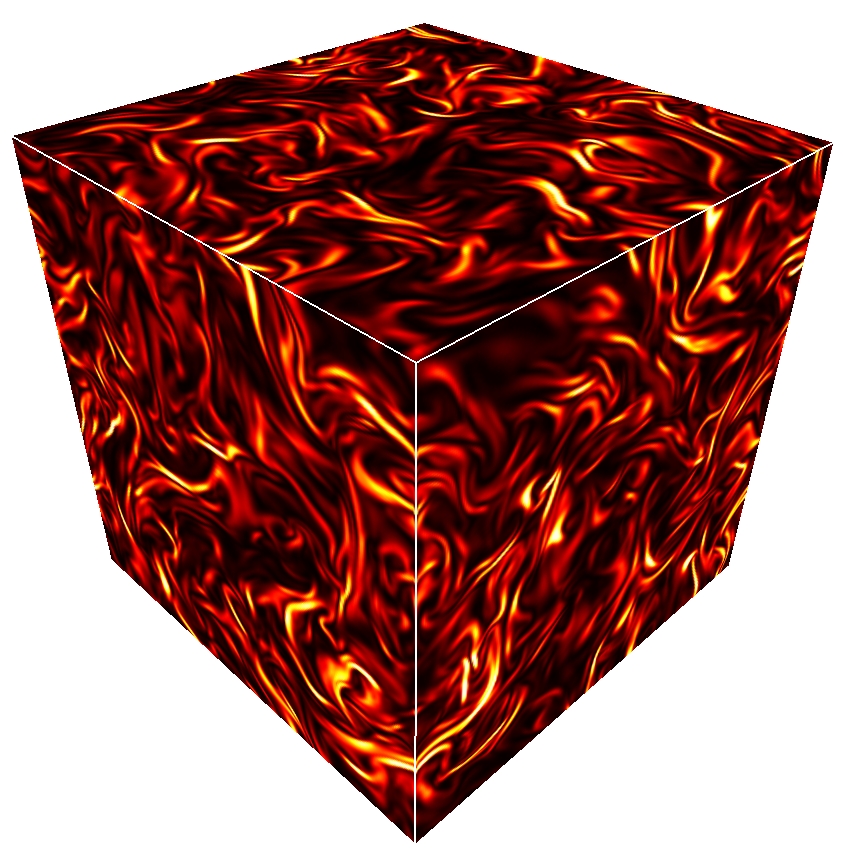}}
        \put(156,-4){\includegraphics[height=120\unitlength]{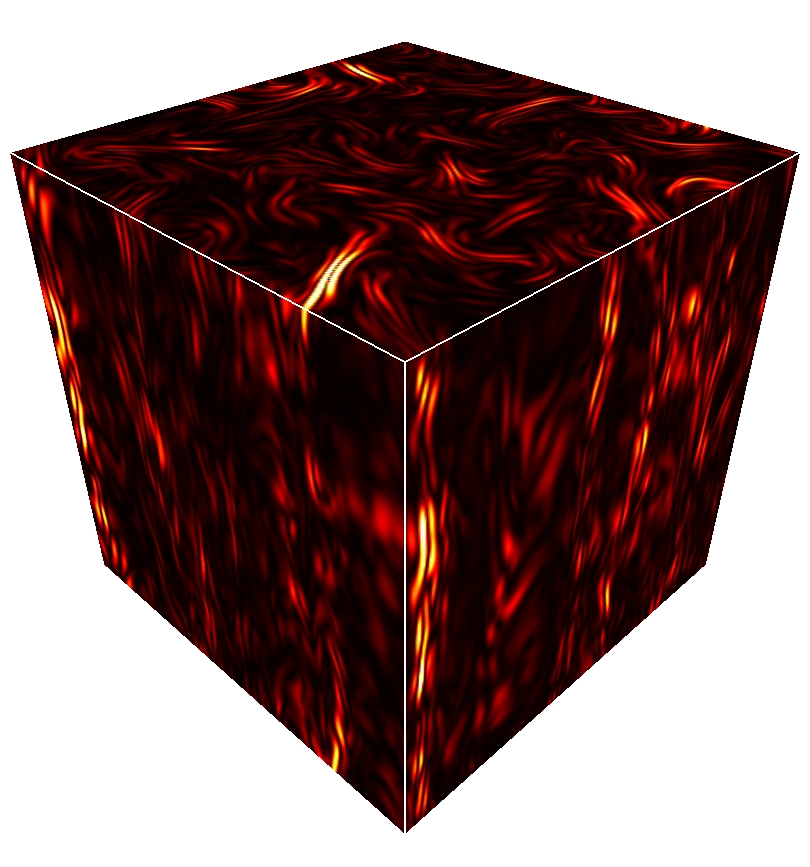}}
	\put(64,8){(a)}
	\put(180,8){(b)}
        \thicklines
	\textcolor{white}{\put(197,90){$\bm{B}_0$}}
	\textcolor{white}{\put(211.5,90){$\bm{\Omega}$}}
        \textcolor{white}{\put(205,84){\vector(0,1){15}}}
	\textcolor{white}{\put(75,90){$\bm{B}_0$}}
        \textcolor{white}{\put(84,84){\vector(0,1){15}}}
        \put(-10,3.5){\includegraphics[height=50\unitlength]{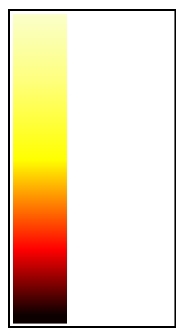}}
	\put(6,8){\small{$0$}}
	\put(2,45){\small{$j_{\textrm{m}}^2$}}
\end{picture}
\caption[]{(Color online) Visualizations of the square current density $\bm{j}^2$ at $t^*=5$. $j_{\textrm{m}}^2$ is about $25\%$ of the maximum value in the numerical domain. (a) Non-rotating MHD case. (b) Rotating MHD case with $\Lambda\approx0.5$.}
\label{fig:visu_rot_mhd_hrm}
\end{figure}
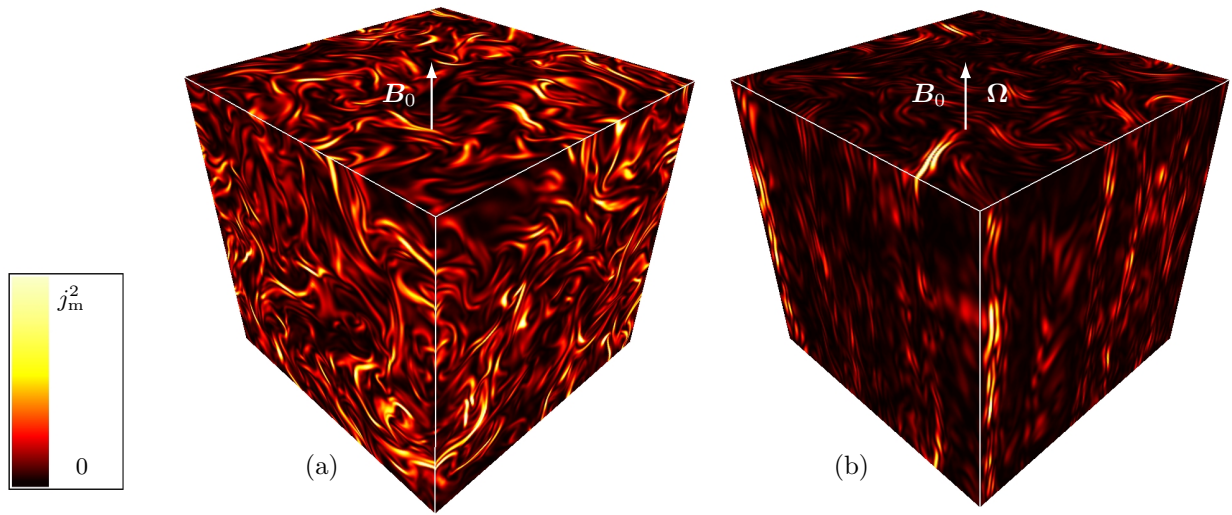

We now focus on the nonlinear turbulence regime of initially homogeneous isotropic developed turbulence submitted to both background rotation and a uniform steady magnetic field.
In that case, the growth of magnetic fluctuations is due to the deformation of the uniform magnetic field lines by the velocity fluctuations, leading to the generation of Alfv\'en waves.
In all the runs, the magnetic fluctuations are initially zero.
We perform different anisotropic simulations to understand the effect of rotation on the growth of magnetic fluctuations.
The main parameters are gathered in Table \ref{tab:param1}.
First, the value of the magnetic diffusivity is fixed so that the initial magnetic Reynolds number based on the integral scale is about $200$, the magnetic Prandtl number being equal to $1$.
The value of the imposed magnetic field corresponds to an interaction parameter of about $12$.
We consider three different rotation rates (Runs $1$ to $3$ in Table \ref{tab:param1}), so that the Rossby number varies from $0.6$ to $0.04$.
For reference, we also perform an isotropic hydrodynamic simulation (run ISO in Table \ref{tab:param1}) and a magnetohydrodynamic simulation without rotation but with an imposed magnetic field (run MHD in Table \ref{tab:param1}).
At time $t=0$, the magnetic field $\bm{B}_0$ and the rotation $\bm{\Omega}$ are applied. Note that the following results are qualitatively unchanged if the rotation is introduced before the magnetic field such that anisotropy due to the Coriolis force is already present when the magnetic field is applied. However, the effect of the Coriolis force on pre-existing magnetic fluctuations is a different matter and will be considered in future works.
In the following, the time is non-dimensionalized by the initial eddy turnover time, $t^*=tu_0/l_0$.
\begin{table}
\begin{center}
\caption{Initial  values of the parameters for DNS computations presented in section \ref{sec:induc}.}
\begin{tabular}{lcccccccc}
\hline
Run & $\nu=\eta$ & $\Omega$ & $\bm{B}_0$ & $R_M$ & $Ro$ & $N$ & $\Lambda$ & $\mathcal{L}$ \\
\hline
ISO & $0.0025$ & $0$ & $0$ & - & - & - & - & - \\
MHD & 0.0025 & $0$ & 0.2 & 200 & - & 12 & $\infty$ & $\infty$ \\
1 & 0.0025 & $1$ & 0.2 & 200 & $0.6$ & 12 & $8$ & $0.16$ \\
2 & 0.0025 & $4$ & 0.2 & 200 & $0.15$ & 12 & $2$ & $0.04$ \\
3 & 0.0025 & $16$ & 0.2 & 200 & $0.04$ & 12 & $0.5$ & $0.01$ \\
\hline
\end{tabular}
\label{tab:param1}
\end{center}
\end{table}

\subsection{Characterization in physical space}
\label{sec:visus}
First, we propose a quick characterization of the flow in physical space. Visualizations of the current density at time $t^*=5$ are gathered in figure \ref{fig:visu_rot_mhd_hrm}.
The non-rotating results are on the left whereas the rotating results are on the right.
Without rotation, the current density is clearly concentrated in sheet-like structures randomly orientated.
Even if the interaction parameter is relatively large, we do not observe well-defined anisotropy since the flow is still in a strong turbulence regime (\textit{i.e.} $B_0\approx u_0$).
In the rotating case ($\Lambda\approx0.5$ presented in figure \ref{fig:visu_rot_mhd_hrm}(b)), the current sheets are clearly aligned with the rotation axis.
Note in addition, that the current sheets are thinner in the rotating case than in the non-rotating case.
Concerning the velocity field in the rotating case (not shown), the visualizations are very similar to what is observed in hydrodynamic rotating turbulence, since the imposed magnetic field is weak: vortices are elongated along the rotation axis.

\begin{figure}
\unitlength 0.4mm
\begin{picture}(250,130)
        \put(12,-5){\includegraphics[height=130\unitlength]{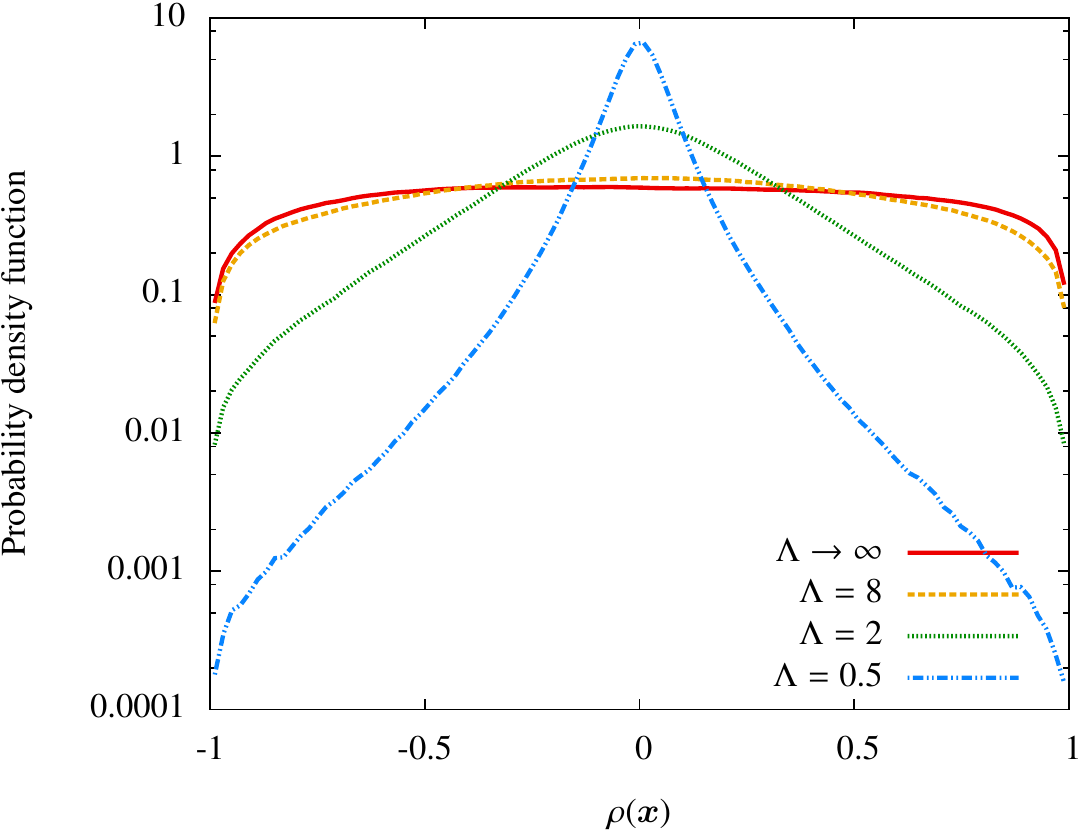}}
        \put(225,-5){\includegraphics[height=130\unitlength]{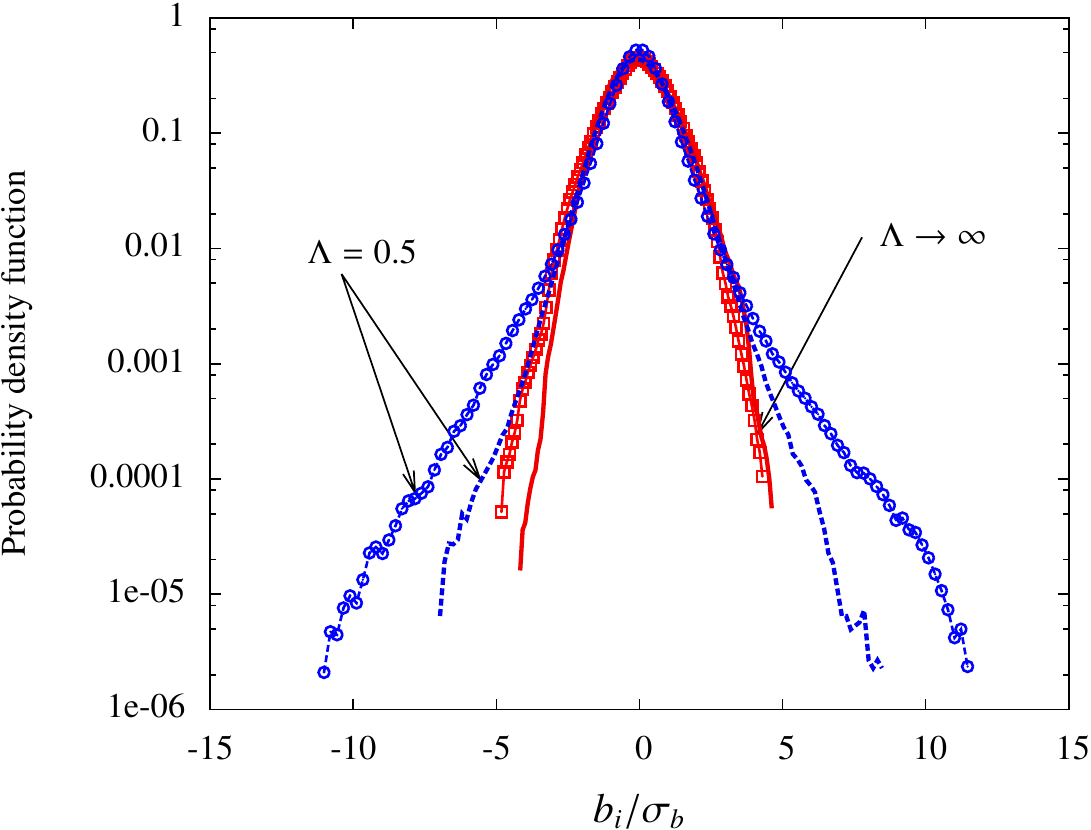}}
	\put(48,110){(a)}
	\put(260,110){(b)}
\end{picture}
\caption[]{(Color online) (a) Normalized pdf of the correlation coefficient $\rho(\bm{x})$ between $\bm{u}$ and $\bm{b}$ at $t^*=5$. (b) Normalized pdf of $\bm{b}/\sigma_b$, where $\sigma_b$ is the variance of $\bm{b}$, at $t^*=5$. The symbols correspond to horizontal components whereas solid lines correspond to vertical components.}
\label{fig:pdf}
\end{figure}

The correlation between the velocity and magnetic field fluctuations can be measured by the cross-correlation coefficient defined by equation \eqref{eq:corrub}. This quantity has already been used for the linear simulations of magneto-inertial waves presented in section \ref{sec:linear}.
In the non-rotating case, more detailed results about this quantity in the context of MHD turbulence can be found in \cite{bigo08}.
First, figure \ref{fig:pdf}(a) presents the probability density function of $\rho(\bm{x})$ at the end of the computation (\textit{i.e.} for $t^*=5$).
Independently of the rotation rate, the pdf are centered around zero (in fact, in the non-rotating case, the pdf is slightly shifted towards negative values, see Bigot \textit{et al.} 2008 for more details).
The extremal values $\rho=-1$ and $\rho=1$ correspond to downward and upward propagating Alfv\'en waves. Note that, in contrast with linear prediction, we do not observe a dominance of Alfv\'enic fluctuations. However, in the strong turbulence regime we consider here, nonlinearities are dominant so that the preferential alignment of Alfv\'en waves is broken. As the rotation increases, one observes that the pdf peaks around zero, indicating that the magnetic fluctuations are mainly oriented perpendicularly to the local velocity fluctuations.
This is consistent with the previous linear analysis. The large scale fluctuations are dominated by inertial waves so that Alfv\'enic fluctuations (characterized by an equipartition between kinetic and magnetic energies, and by $|\rho(\bm{x})|\approx1$) are damped.

The pdf of the vertical (solid lines) and horizontal components (symbols) of the fluctuating magnetic field are gathered on figure \ref{fig:pdf}(b). One can see that the horizontal component of the magnetic field is dominant and more intermittent in rotating cases than in non-rotating case, which is consistent with the visualizations, where very thin current sheets are aligned with the rotation axis.
\subsection{Dynamics}
\label{sec:ehrm}
\begin{figure}[b!]
\unitlength 0.4mm
\begin{picture}(250,140)
        \put(20,-5){\includegraphics[height=140\unitlength]{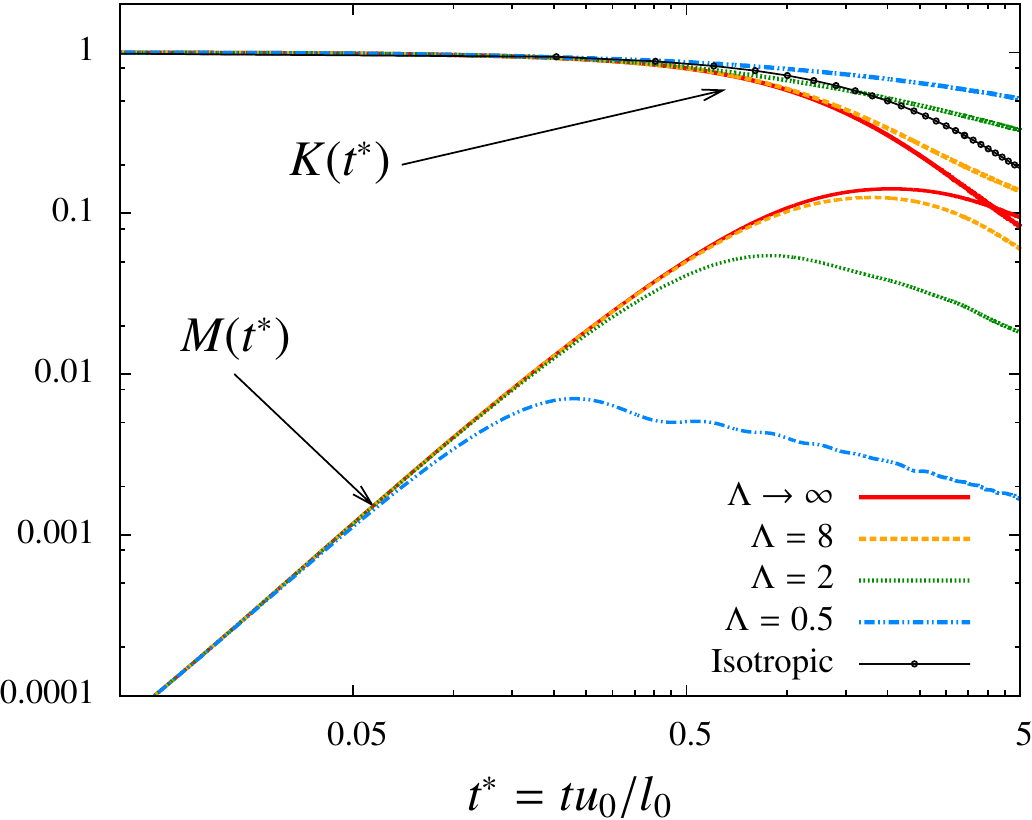}}
        \put(230,-5){\includegraphics[height=140\unitlength]{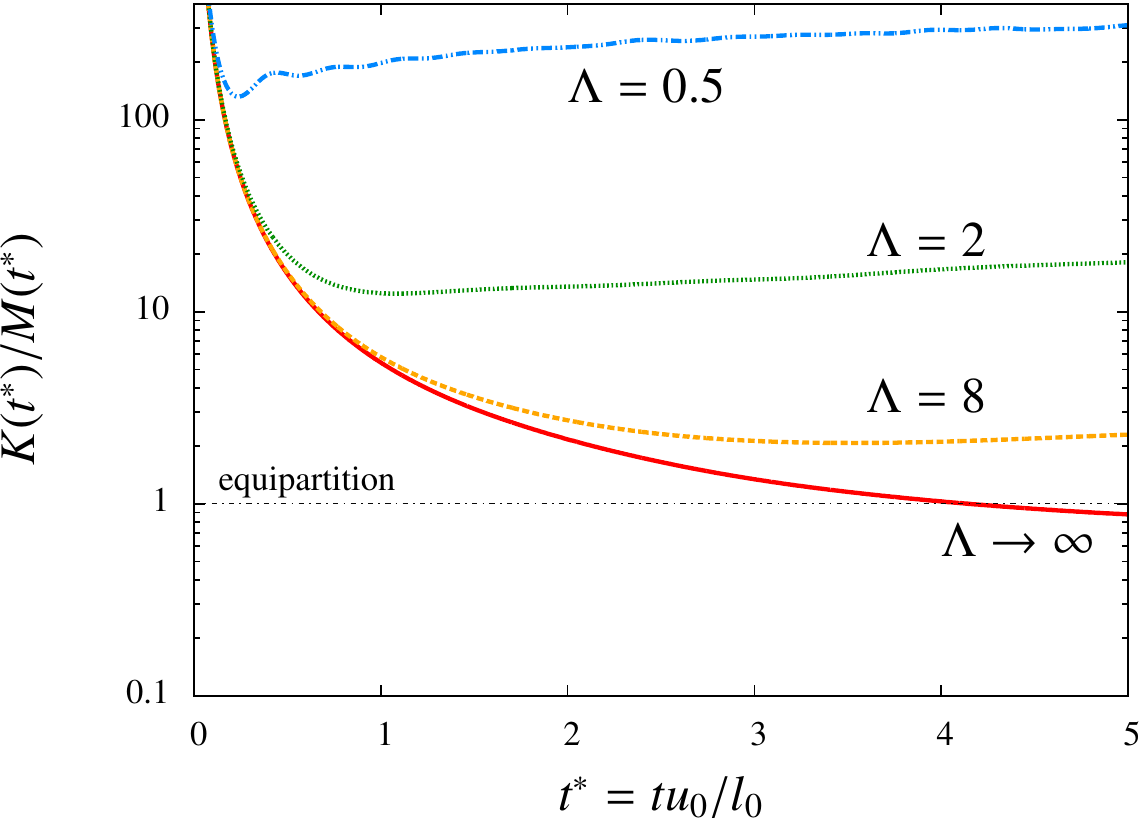}}
	\put(47,20){(a)}
	\put(264,20){(b)}
\end{picture}
\caption[]{(Color online) (a) Kinetic energy $K$ and magnetic energy $M$ versus dimensionless time $t^*=tu_0/l_0$. (b) Alfv\'en ratio $K/M$.}
\label{fig:ehighrm}
\end{figure}

Figure \ref{fig:ehighrm}(a) presents the time evolution of the magnetic (bottom of the figure) and kinetic energies (top of the figure) for different values of the Elsasser number.
In addition, figure \ref{fig:ehighrm}(b) represents the ratio between the kinetic energy and the magnetic energy, also called Alfv\'en ratio.
In the wave turbulence regime (\textit{i.e.} $B_0\gg u_0$), this ratio is expected to be equal to one, since Alfv\'en waves are dominant and characterized by an equipartition between kinetic and magnetic energies.
However, non-rotating strong MHD turbulence submitted to a uniform magnetic field is characterized by an Alfv\'en ratio slightly smaller than one, indicating the presence of non-Alfv\'enic fluctuations (see \textit{e.g.} Bigot \textit{et al.} 2008).
In the non-rotating case (see the $\Lambda\rightarrow\infty$ curve on figure \ref{fig:ehighrm}(b)), this quasi-equipartition state is indeed observed at the end of the simulation (\textit{i.e.} $t^*>4$), with a slight dominance of magnetic energy.
However, as the Elsasser number decreases, \textit{i.e.} as the rotation rate increases whereas the imposed magnetic field remains constant, the magnetic fluctuations are strongly damped whereas the kinetic energy decrease is slower than in non-rotating cases. Thus, the Alfv\'en ratio increases with the rotation rate, indicating that Alfv\'enic fluctuations are dominated by another phenomenon, namely inertial waves. This is in accordance with the linear predictions of section \ref{sec:linear}, and the similarities between figures \ref{fig:linear_results}(a) and \ref{fig:ehighrm}(a) show that waves play an important part in the dynamics.
The line with symbols on figure \ref{fig:ehighrm}(a) corresponds to the isotropic computation (\textit{i.e.} $B_0=0$ and $\Omega=0$) obtained from the same initial condition. Without rotation (\textit{i.e.} $\Lambda\rightarrow\infty$), the kinetic energy decays faster in the magnetized case than in the hydrodynamic case due to the additional ohmic dissipation.
As the rotation rate increases, the kinetic energy decay rate is slower, which is a well-known result already observed for non-magnetized rotating turbulence \citep{jacq90}.
However, the damping of the magnetic energy due to background rotation is an original observation.
\begin{figure}
\unitlength 0.4mm
\begin{picture}(250,140)
        \put(10,-5){\includegraphics[height=140\unitlength]{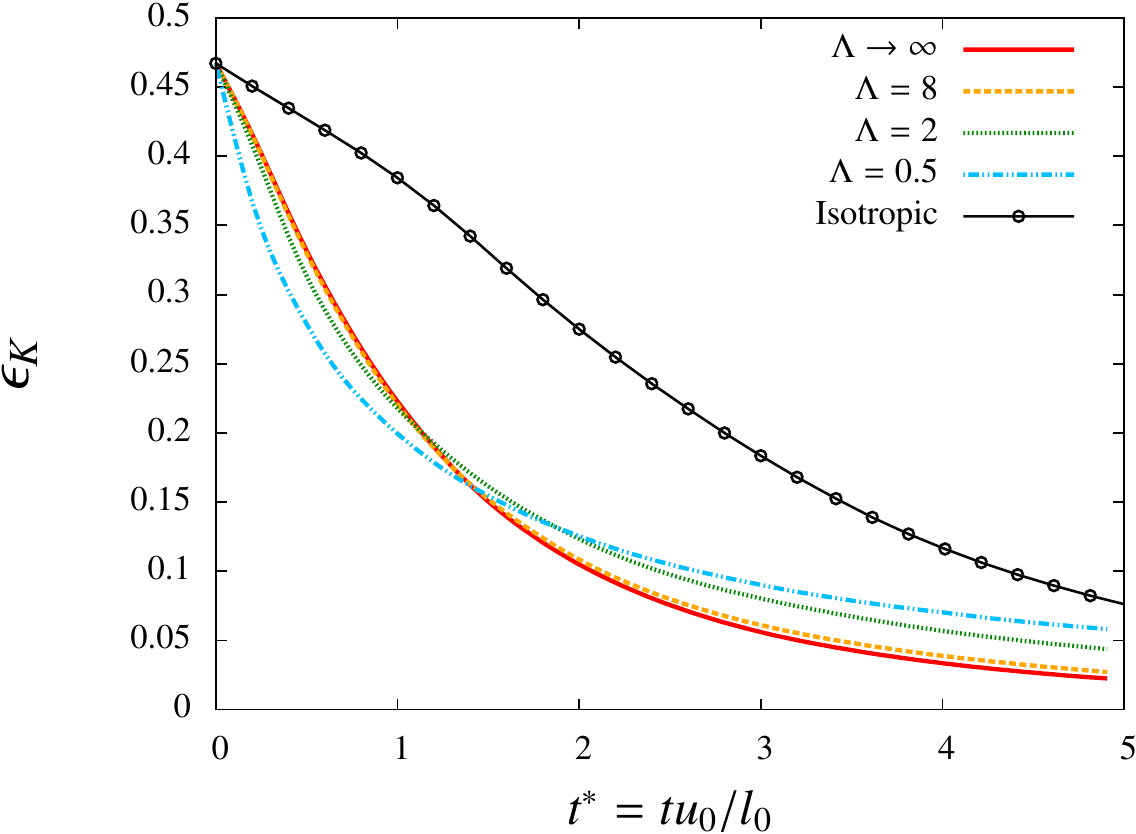}}
        \put(225,-5){\includegraphics[height=140\unitlength]{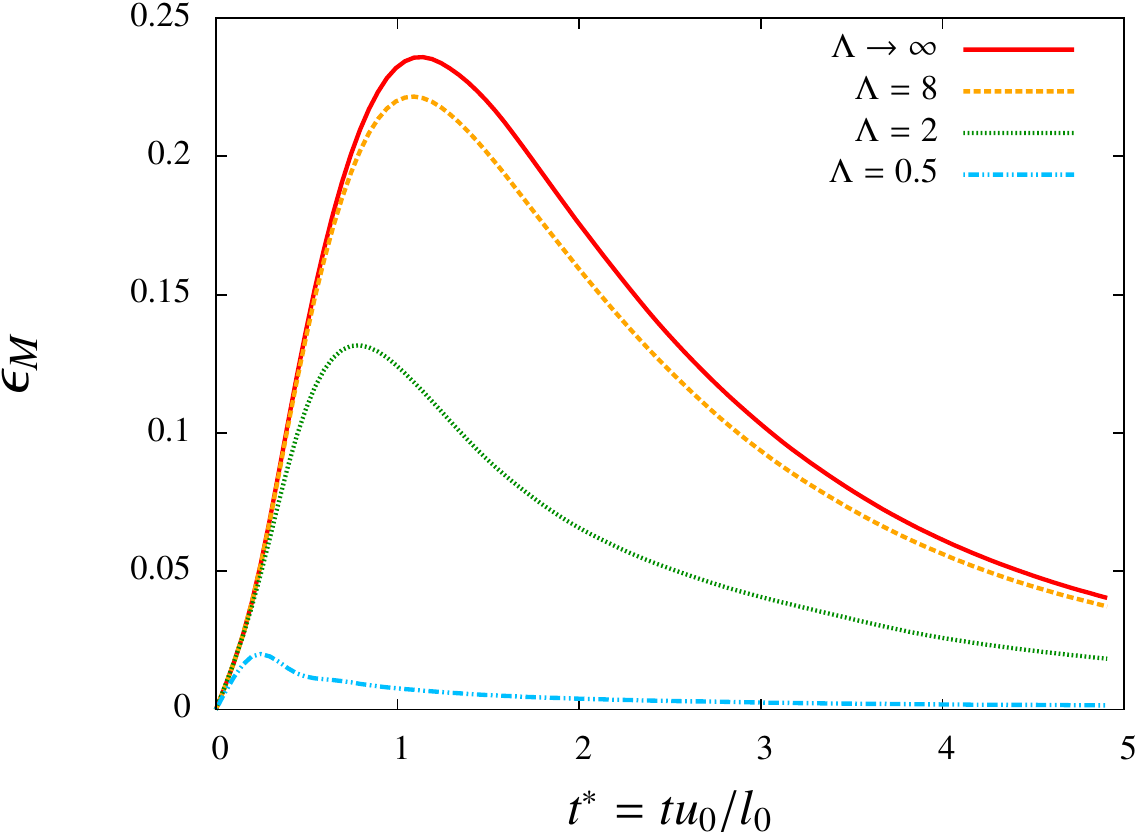}}
	\put(55,122){(a)}
	\put(264,120){(b)}
\end{picture}
\caption[]{(Color online) (a) Kinetic energy dissipation rate $\epsilon_K$ versus time. (b) Magnetic energy dissipation rate $\epsilon_K$ versus time.}
\label{fig:eps}
\end{figure}

\begin{figure}[b!]
\unitlength 0.4mm
\begin{picture}(250,260)
        \put(10,125){\includegraphics[height=130\unitlength]{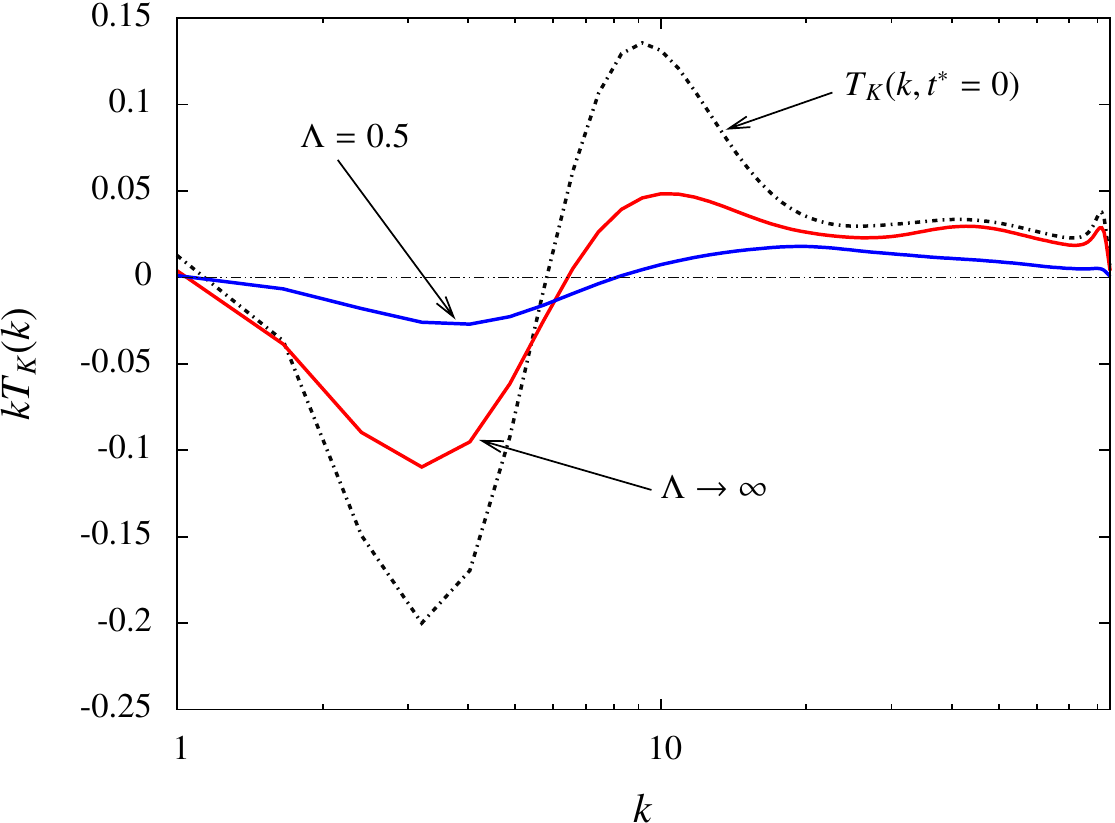}}
        \put(220,125){\includegraphics[height=130\unitlength]{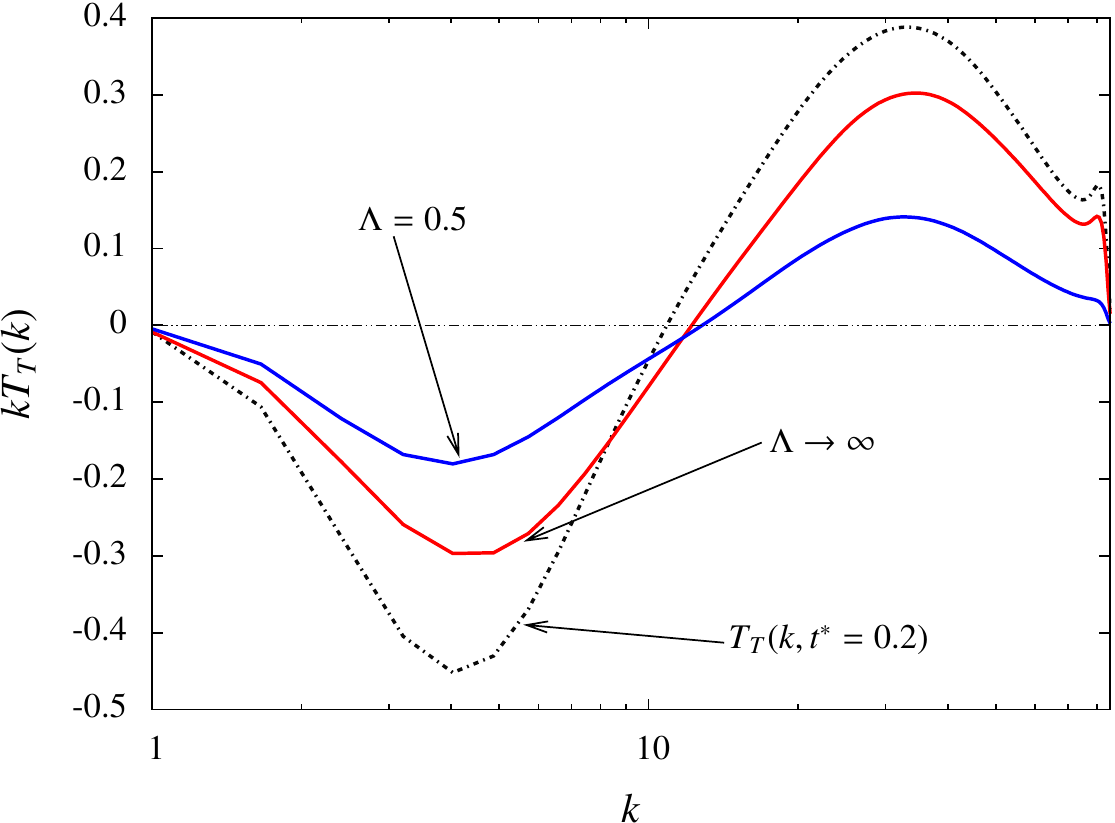}}
        \put(110,-5){\includegraphics[height=130\unitlength]{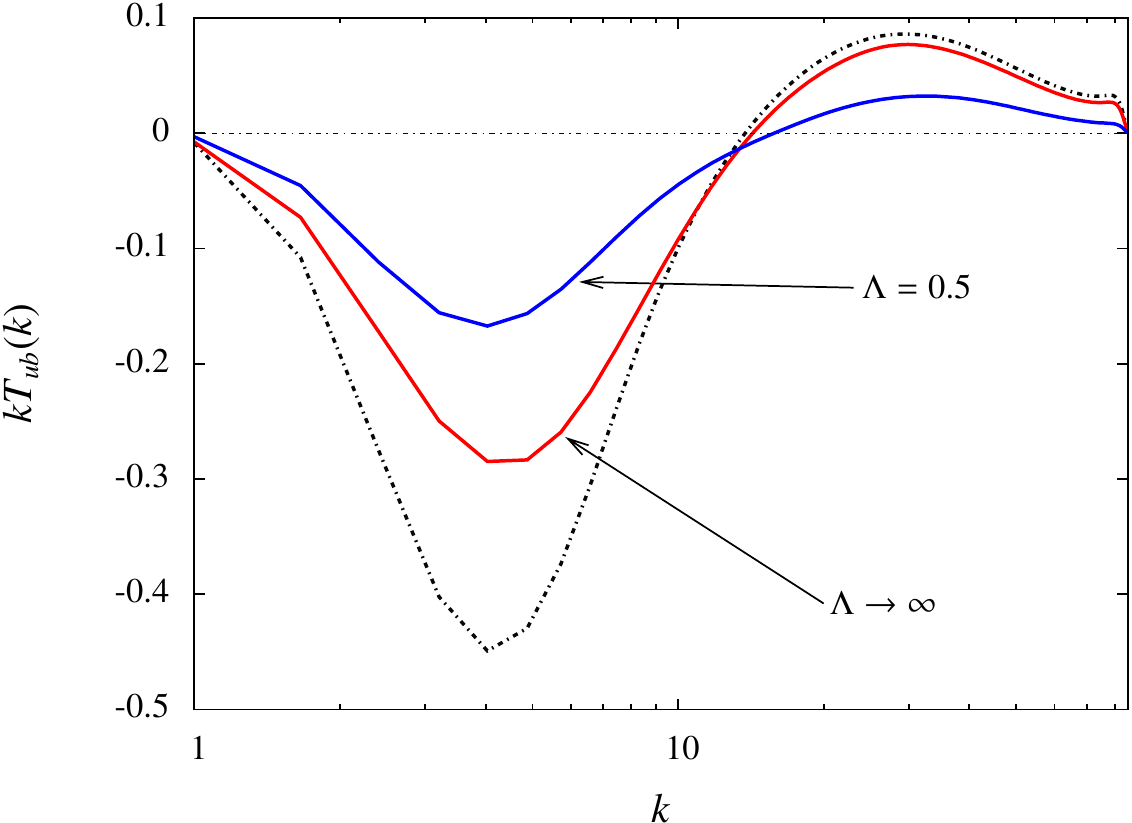}}
	\put(43,240){(a)}
	\put(255,240){(b)}
	\put(145,110){(c)}
\end{picture}
\caption[]{(Color online) (a) Kinetic energy transfer spectrum $T_K(k)$. The dashed thick line corresponds to the initial transfer spectrum at $t^*=0$. Solid lines correspond to transfer spectra at $t^*=2$. (b) Total energy transfer spectrum $T_T(k)$. Dashed lines correspond to transfer spectra at $t^*=0.2$ whereas solid lines correspond to $t^*=2$. Given that the figures are plotted in semi-log scale, the transfers are multiplied by $k$ to enhance the fact that their integrals are zero.
(c) Spectra of $\bm{b}{\boldsymbol{\cdot}}\bm{q}$. The dashed line corresponds to the spectrum at $t^*=0.2$ whereas solid lines correspond to spectra at $t^*=2$.}
\label{fig:trnl}
\end{figure}
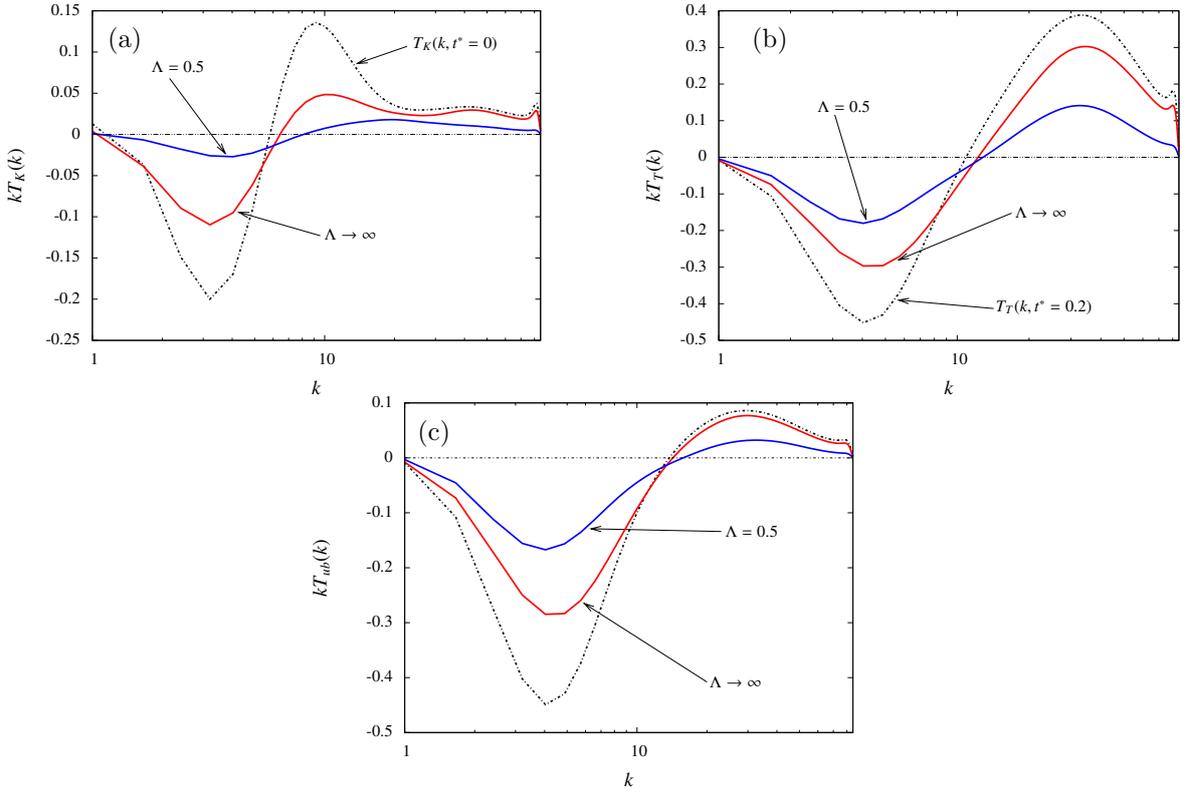

We define the kinetic dissipation rate $\epsilon_K$ and the magnetic dissipation rate $\epsilon_M$ as
\begin{align}
\epsilon_K & =\nu\left<\bm{\omega}^2\right> \\
\epsilon_M & =\eta\left<\bm{j}^2\right> \ .
\end{align}
$\epsilon_K$ and $\epsilon_M$ are plotted in figures \ref{fig:eps}(a) and (b) respectively.
The kinetic dissipation rate is always decreasing, since we compute MHD flows from developed turbulence already containing small-scale fluctuations, and is always smaller than its isotropic value (line with symbols in fig.\ref{fig:eps}(a)). It seems inconsistent to observe a smaller value of the kinetic dissipation between isotropic and MHD turbulence, since the latter contains additional dissipative effects. Note however that in the present MHD case without initial magnetic fluctuations, some of the initial kinetic energy is transferred to the magnetic energy, thus reducing the amplitude of the kinetic motion and dissipation. Note in addition that the sum of both dissipation rates $\epsilon_K$ and $\epsilon_M$ (not shown) is greater than the isotropic kinetic dissipation rate.
The initial effect of rotation is, as in non-magnetized cases \citep{camb01,mori01}, to decrease the kinetic energy dissipation.
However, at larger times (\textit{i.e.} for $t^*>2$), $\epsilon_K$ is greater for rotating cases than for non-rotating ones. This result deserves some explanations. The well-known effect of the Coriolis force is to decrease kinetic energy transfer (see figure \ref{fig:trnl}(a) below). This implies a reduction of $\epsilon_K$ at the beginning of the simulation. However, as time increases, the decay of $\epsilon_K$ is also reduced by rotation (since the kinetic energy is less dissipated), so that it can exceed its non-rotating value.

Concerning the magnetic field, we observe an initial increase of the magnetic dissipation rate due to the generation of small-scale magnetic fluctuations. After reaching a maximum, $\epsilon_M$ decreases. As rotation is introduced, the small-scale development of the magnetic fluctuations is accelerated (since the maximum of $\epsilon_M$ occurs at earlier times) but the maximum of the magnetic dissipation is substantially reduced. The reduction of $\epsilon_M$ with the rotation rate compensates the increase of $\epsilon_K$ at large times, so that the whole dissipation rate $\epsilon_K+\epsilon_M$ always decreases with the Rossby number.

Let us now discuss the transfers of both kinetic and magnetic energies.
In the equation \eqref{eq:momentum} governing $\bm{u}$, in addition to the classical nonlinear advective term $\bm{s}^{(u)}=\bm{u}\times\bm{\omega}$, the nonlinear term which reflects the impact of the fluctuating magnetic field is $\bm{s}^{(b)}=\bm{j}\times\bm{b}=({\boldsymbol{\nabla}} \times \bm{b}) \times \bm{b}$.
Similarly, the advective term in the induction equation \eqref{eq:induction} is $\bm{q}={\boldsymbol{\nabla}}\times(\bm{u}\times\bm{b})$, which can be rewritten as
\begin{equation}
q_i = \frac{\partial}{\partial x_j} (u_i b_j - u_j b_i)= -\frac{\partial }{\partial x_j} (b_i u_j) + \frac{\partial u_i}{\partial x_j}b_j
\end{equation}
with implicit summation over repeated indices.
This equation can be recovered using Ricci formulae, but is well known with a complete analogy with the Helmholtz equation for fluctuating vorticity.

The dot product of equation \eqref{eq:momentum} by $\bm{u}$ yields the equation for the kinetic energy $K=\frac12\bm{u}{\boldsymbol{\cdot}}\bm{u}$, or
\begin{equation}
\label{eq:ek}
\frac{\partial K}{\partial t}-\nu\bm{u}{\boldsymbol{\cdot}}\nabla^2\bm{u}=\bm{u}{\boldsymbol{\cdot}}\bm{s}^{(u)}+2\bm{u}{\boldsymbol{\cdot}}(\bm{u}\times\bm{\Omega})+\underbrace{\bm{u}{\boldsymbol{\cdot}}(\bm{j}\times\bm{B}_0)}_{(a)}+\bm{u}{\boldsymbol{\cdot}}\bm{s}^{(b)}-\bm{u}{\boldsymbol{\cdot}}{\boldsymbol{\nabla}} P \ .
\end{equation}
Terms which are obviously zero, such as $\bm{u}{\boldsymbol{\cdot}}\bm{s}^{(u)}$, are retained in equation \eqref{eq:ek} in order to anticipate the origin of ``true'' spectral transfer terms with zero integral.
The dot product of equation \eqref{eq:induction} by $\bm{b}$ yields the equation for the magnetic energy $M=\frac12\bm{b}{\boldsymbol{\cdot}}\bm{b}$, or
\begin{equation}
\label{eq:em}
\frac{\partial M}{\partial t}-\eta\bm{b}{\boldsymbol{\cdot}}\nabla^2\bm{b}=\underbrace{\bm{b}{\boldsymbol{\cdot}}({\boldsymbol{\nabla}}\times(\bm{u}\times\bm{B}_0))}_{(b)}+\bm{b}{\boldsymbol{\cdot}}\bm{q} \ .
\end{equation}
Important cubic terms in the kinetic energy equation are
\begin{equation}
\label{eq:cubick}
\bm{u} {\boldsymbol{\cdot}} \bm{s}^{(b)} = \frac{\partial }{\partial x_i} (u_jb_j b_i)-\frac{\partial u_i}{\partial x_j}b_i b_j - \frac{1}{2} \frac {\partial }{\partial x_i} (b^2 u_i) \ .
\end{equation}
Their counterpart in equation \eqref{eq:em} for $M$ are
\begin{equation}
\label{eq:cubicm}
\bm{b} {\boldsymbol{\cdot}} \bm{q} = \frac{\partial u_i}{\partial x_j} b_i b_j - \frac{1}{2}
\frac{\partial }{\partial x_i}(b^2 u_i) \ .
\end{equation}

Energy conservation is found by taking ensemble averages of equations \eqref{eq:ek} and \eqref{eq:em} and ignoring dissipation terms.
Mean field terms play no role: the second term in the right hand side of equation \eqref{eq:ek} is zero because the Coriolis force produces no energy whereas the terms $(a)$ and $(b)$ in equations \eqref{eq:ek} and \eqref{eq:em} linked to Alfv\'en velocity cancel each others.
Only the cubic terms in equations \eqref{eq:cubick} and \eqref{eq:cubicm} control the energy transfer.
Terms which appear as the gradient of a cubic product are zero due to statistical homogeneity and their spectra really correspond to a true downscale transfer.
In contrast, the term $\langle \partial u_i/\partial x_jb_i b_j\rangle$ is similar to the nonlinear vortex stretching term in the hydrodynamic case, replacing $\bm{b}$ by $\bm{\omega}$.
It is nonzero and does not correspond to a transfer term.
In counterpart, this term appears with opposite sign in
the kinetic equation \eqref{eq:ek} and the magnetic equation \eqref{eq:em},
so that it is a pure `intercomponent' (from magnetic to kinetic)
energy transfer, although not a true `scale-to-scale' energy transfer,
and has no contribution to the equation for total
energy $K+M$.

Figure \ref{fig:trnl}(a) presents the classical spherically averaged transfer spectra $T_K(k)$ derived from the advective term $\bm{u}{\boldsymbol{\cdot}}\bm{s}^{(u)}$ in equation \eqref{eq:ek}.
We plot the value of the transfer multiplied by the wave number so that, in semi-log scale, one clearly sees that the positive area is equal to the negative one.
As for non-magnetized rotating turbulence, the rotation tends to reduce nonlinear transfer from large scales to small scales, thus reducing the overall kinetic energy dissipation (at least initially, see figure \ref{fig:eps}(a)).
Figure \ref{fig:trnl}(b) presents the transfer spectra $T_T(k)$ derived from the sum $\bm{u}{\boldsymbol{\cdot}}\bm{s}^{(b)}+\bm{b}{\boldsymbol{\cdot}}\bm{q}$, which correspond to the spectral transfer in the equation for the total energy.
Again, the effect of the rotation is clearly to reduce nonlinear transfers, thus reducing the total (\textit{i.e.} kinetic plus magnetic) dissipation.
Finally, the advection of the magnetic fluctuations by the velocity field is characterized by the term $\bm{b}{\boldsymbol{\cdot}}\bm{q}$.
This term is not a transfer (its integral over $k$ is non zero) but is however also reduced by the rotation, as shown on figure \ref{fig:trnl}(c).

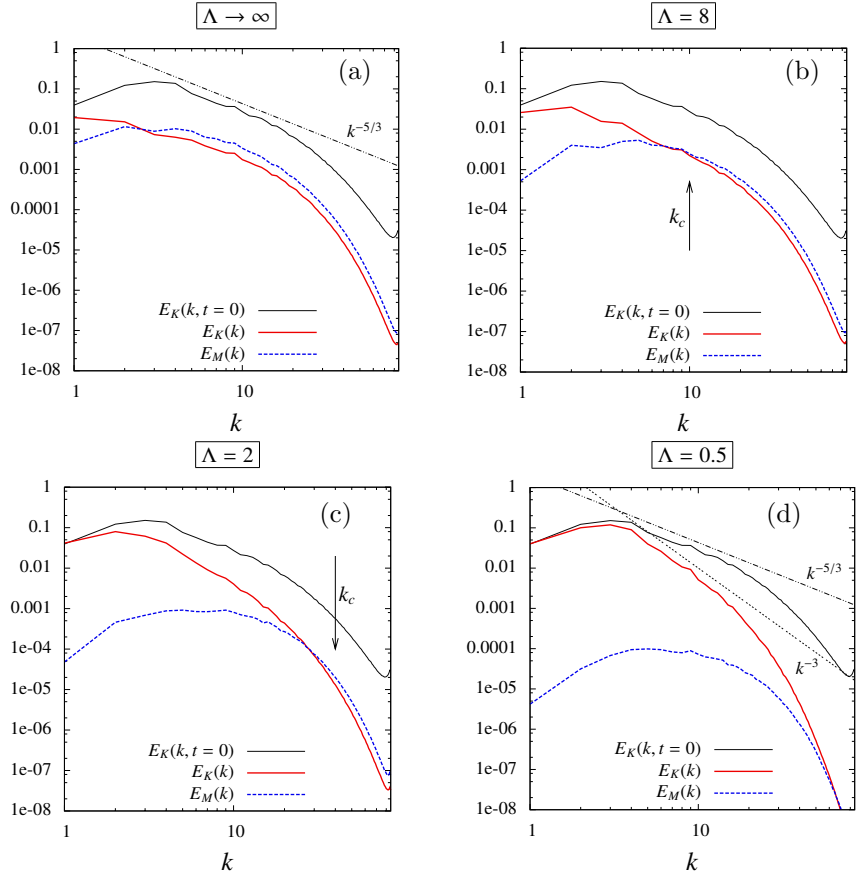
\begin{figure}[t!]
\unitlength 0.6mm
\begin{picture}(250,190)
        \put(40,100){\includegraphics[height=95\unitlength]{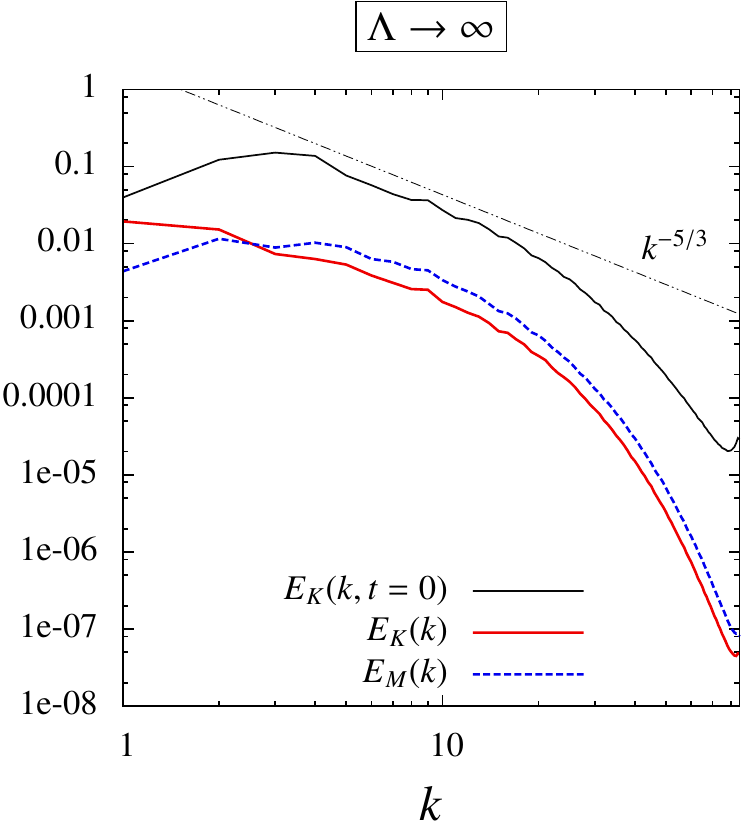}}
        \put(140,100){\includegraphics[height=95\unitlength]{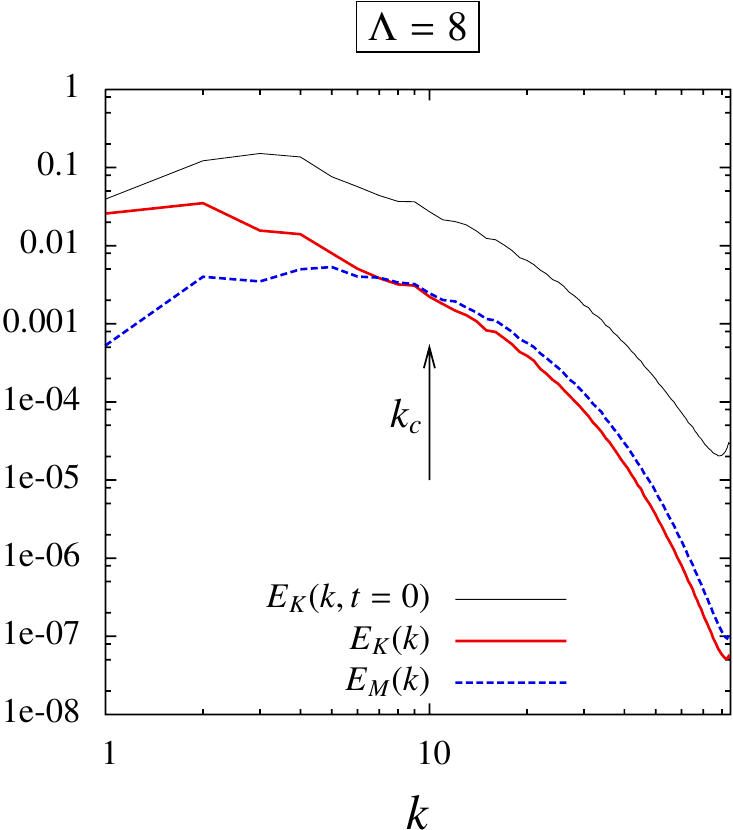}}
        \put(40,3){\includegraphics[height=95\unitlength]{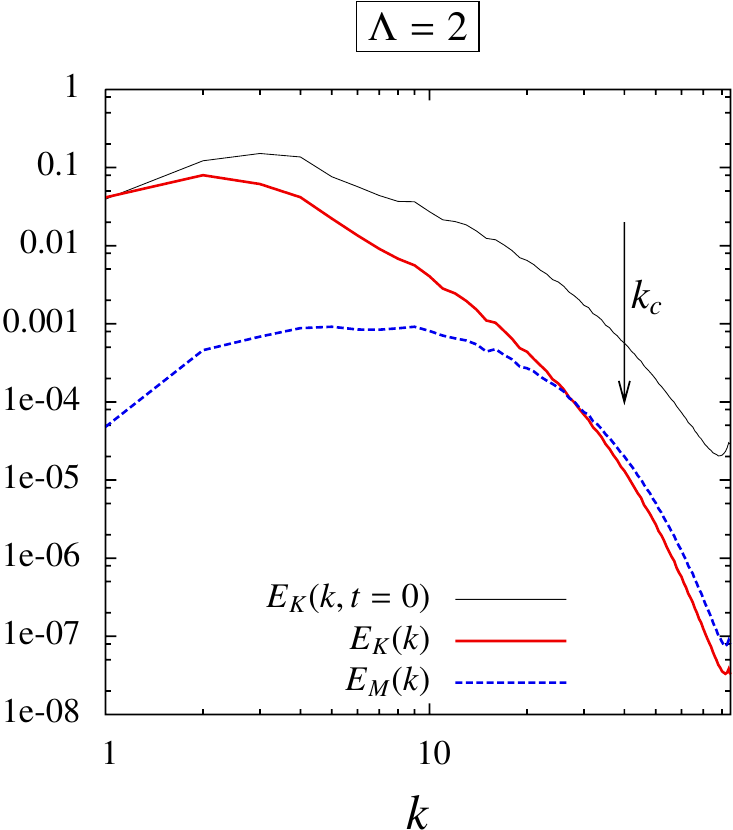}}
        \put(140,3){\includegraphics[height=95\unitlength]{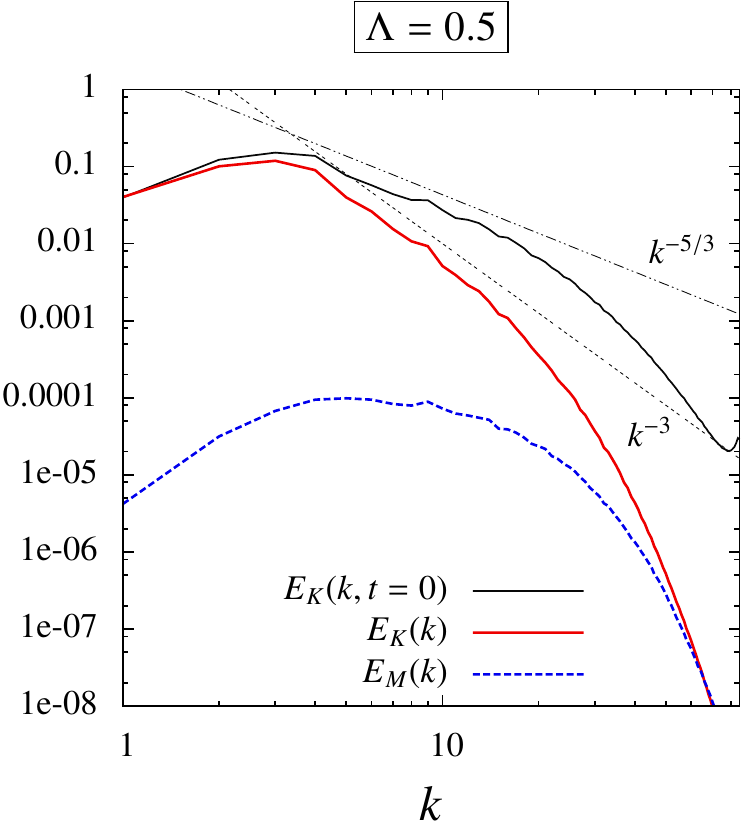}}
	\put(112,178){(a)}
	\put(210,178){(b)}
	\put(108,80){(c)}
	\put(206,80){(d)}
\end{picture}
\caption[]{(Color online) Kinetic energy spectra $E_K(k)$ and magnetic energy spectra $E_M(k)$ at $t^*=5$. Vertical arrows corresponds to the cut-off wave number $k_c=2\Omega/B_0$. The $k^{-5/3}$ and $k^{-3}$ slopes are shown for comparison with classical scalings in hydrodynamic isotropic and rotating turbulence (see for example Sagaut and Cambon 2008). (a) $\Lambda\rightarrow\infty$. (b) $\Lambda=8$. (c) $\Lambda=2$. (d) $\Lambda=0.5$.}
\label{fig:comp_spec_1}
\end{figure}

Spherically averaged energy spectra are represented in figure \ref{fig:comp_spec_1}, for four different Elsasser numbers from $\infty$ (non-rotating case) to $0.5$.
All the spectra are plotted at the same dimensionless time $t^*\approx5$.
The magnetic energy spectra $E_M(k)$ correspond to the dotted blue lines whereas the kinetic energy spectra correspond to the solid red lines.
In each graph, the initial kinetic energy spectrum is presented for reference as a thin solid line. We retrieve the global reduction of the magnetic energy while increasing the rotation rate, as already observed in figure \ref{fig:ehighrm}.
Without rotation ($\Lambda\rightarrow\infty$, figure \ref{fig:comp_spec_1}(a)), the kinetic/magnetic quasi-equipartition of energy is observed at all scales of the flow, with a slight excess of magnetic energy with respect to the kinetic one for inertial and dissipative scales.
At moderate Elsasser number ($\Lambda\approx8$, figure \ref{fig:comp_spec_1}(b)), the rotation affects predominantly the large scales of the flow, \textit{i.e.} the magnetic fluctuations are attenuated at small wave numbers, whereas the large wave numbers are still equipartitioned, with an Alfv\'en ratio about unity.
When rotation increases (figures \ref{fig:comp_spec_1}(c) and (d)), the magnetic energy is damped over all scales of the flow, whereas the kinetic energy spectrum at small wave numbers recovers the initial isotropic level, which is consistent with the decrease of the kinetic direct cascade.
This scale-dependent attenuation of the magnetic energy is directly linked to the competition between Alfv\'enic fluctuations and inertial waves due to rotation, as shown by the previous linear analysis.

It is important to note that the dispersion properties of inertial waves are very different from Alfv\'en waves.
At a given scale $k$, the maximum frequency of Alfv\'en waves is $B_0k$ whereas the maximum frequency of inertial waves is $2\Omega$.
By equating these two frequencies, one can therefore estimate a cut-off wave number $k_c=2\Omega/B_0$.
The value of $k_c$ is indicated by a vertical arrow in figures \ref{fig:comp_spec_1}(b) and (c), which clearly separates two spectral domains in wavespace. In the spectral domain such that $k<k_c$, inertial waves are the fastest phenomenon, so that the energy transfers from kinetic to magnetic fluctuations due to Alfv\'en waves are damped; whereas in the region $k>k_c$, Alfv\'enic fluctuations are faster than inertial waves and an equipartition of energy is still observed.
%
%
\subsection{Spectral anisotropy}
This section is devoted to the characterization in spectral space of the anisotropy of rotating MHD turbulence at high magnetic Reynolds number.
The anisotropy of the flow is described using different statistical quantities.

\subsubsection{Shebalin angles}

The Shebalin angle $\theta_Q$ is often used in the study of MHD turbulence \citep{voro05,sheb83} and is defined, for any spectral quantity $\hat{\bm{Q}}(\bm{k},t)$, by
\begin{equation}
\tan^2\theta_Q=\frac{\sum_{\bm{k}}k_{\mathrm{h}}^2|\hat{\bm{Q}}(\bm{k},t)|^2}{\sum_{\bm{k}}k_z^2|\hat{\bm{Q}}(\bm{k},t)|^2}
\label{eq:sheb}
\end{equation}
where $k_{\mathrm{h}}=\sqrt{k_x^2+k_y^2}$ is the horizontal component of the wave vector $\bm{k}$, whereas $k_z$ is the vertical component (aligned with the rotation axis).
For an isotropic quantity, $\theta_Q \approx 55^{\textrm{o}}$.
A Shebalin angle of $90^{\textrm{o}}$ is characteristic of a quantity independent of the vertical direction. This quantity and similar indicators are used in \cite{favier10} for quasistatic MHD turbulence.

\begin{figure}[t!]
\unitlength 0.4mm
\begin{picture}(250,140)
        \put(5,-5){\includegraphics[height=140\unitlength]{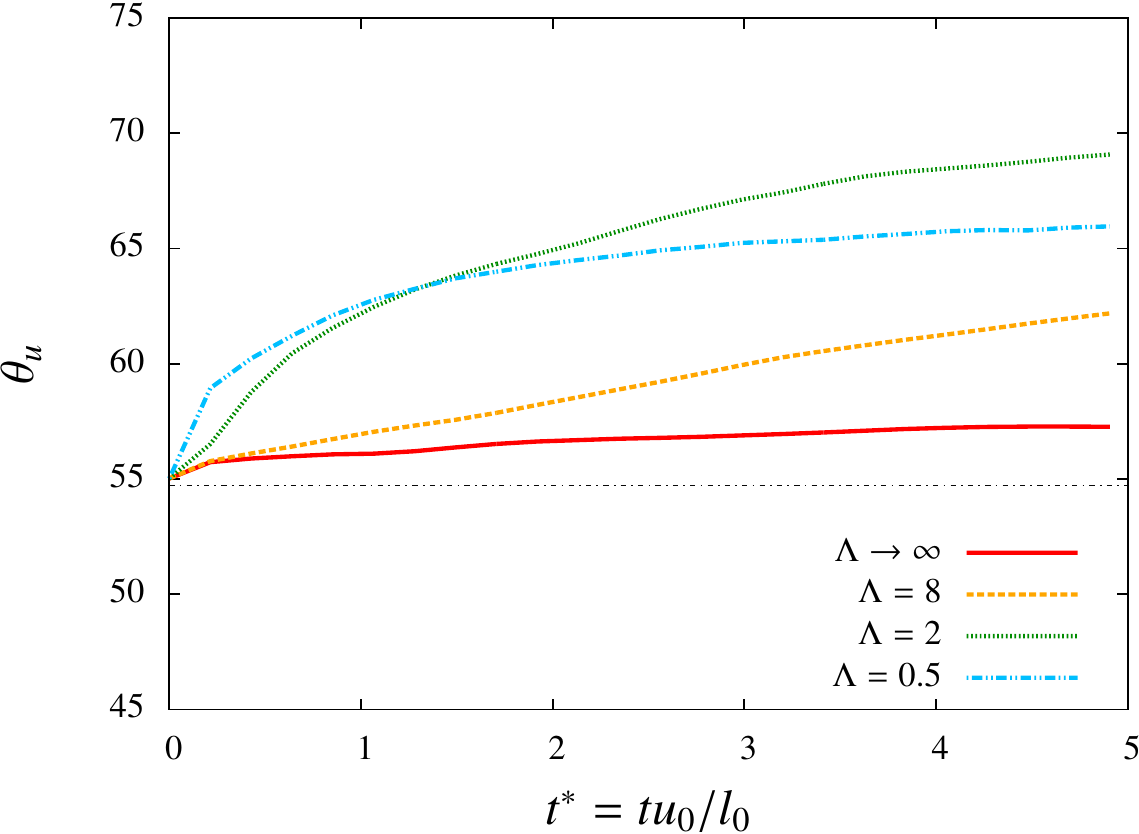}}
        \put(220,-5){\includegraphics[height=140\unitlength]{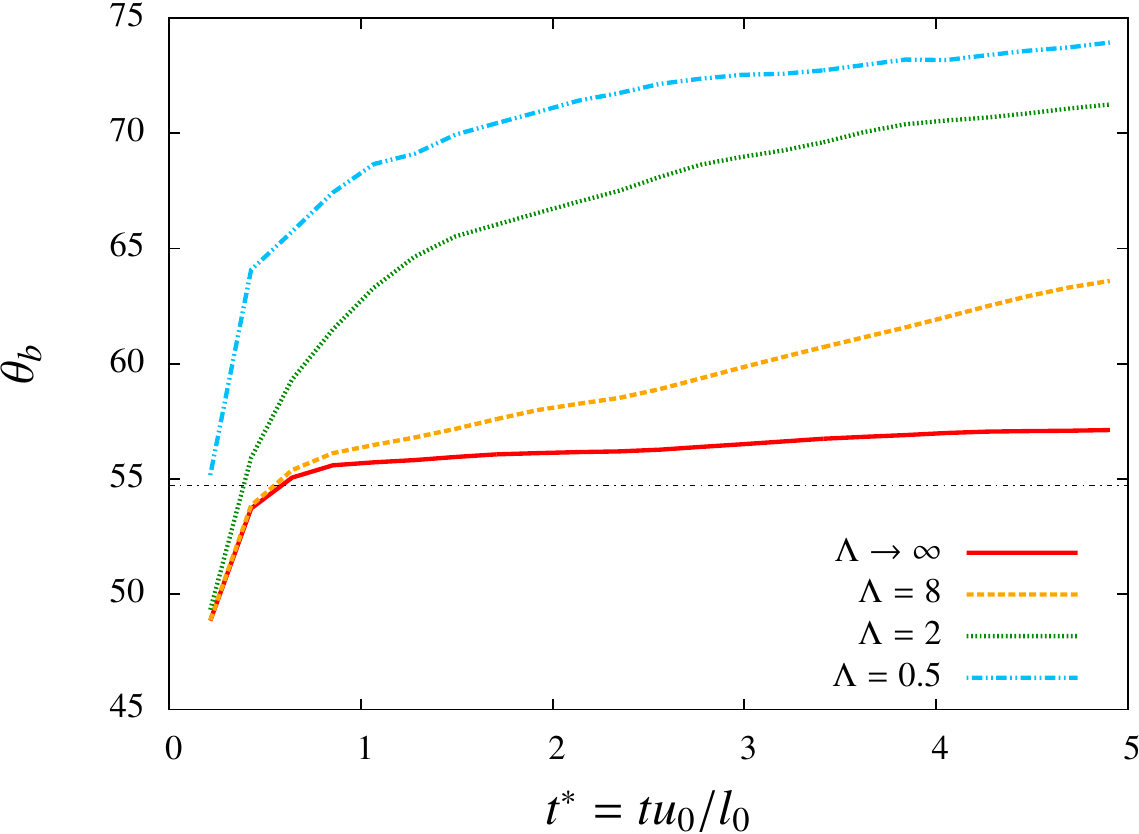}}
	\put(45,19){(a)}
	\put(256,19){(b)}
\end{picture}
\caption[]{(Color online) Shebalin angles versus dimensionless time $t^*=tu_0/l_0$. (a) $\theta_u(t^*)$. (b) $\theta_b(t^*)$.}
\label{fig:sheb_hrm}
\end{figure}

The Shebalin angles of the kinetic and magnetic energy spectra $E_K(k,\theta)$ and $E_M(k,\theta)$, $\theta_u$ and $\theta_b$, are gathered in figures \ref{fig:sheb_hrm}(a) and (b) respectively.
Without rotation, both velocity and fluctuating magnetic fields are nearly isotropic with a Shebalin angle slightly greater than the isotropic value. This reflects a slight accumulation of kinetic and magnetic energies on equatorial modes such that $\bm{k}\perp\bm{B}_0$, due to the damping of nonlinear transfers along the imposed magnetic field \citep{ough94,alex07}. Note that, in the non-rotating case, the anisotropy is moderate since the interaction parameter is relatively small compared to previous studies \citep{bigo08}.
However, as the rotation increases (\textit{i.e.} as the Elsasser number decreases), both fields become strongly anisotropic since the two Shebalin angles increase.
A Shebalin angle close to $90^{\textrm{o}}$ is characteristic of a flow which tends to be invariant in the vertical direction, \textit{i.e.} in which the vertical gradients $\partial/\partial z$ decrease strongly, or, equivalently, the spectral energy concentrates in equatorial modes such that $\bm{k}\perp\bm{\Omega}$.
Unlike the monotonous evolution of $\theta_b$ with $\Lambda$, after a first increase with $\Lambda$ from $\infty$ to $0.5$, $\theta_u$ decreases at very low Rossby numbers, as illustrated by the curves at $\Lambda=2$ and $\Lambda=0.5$ on figure \ref{fig:sheb_hrm}(a).
This effect has already been observed in hydrodynamic rotating turbulence at very low Rossby number \citep{CAMBON-MANSOUR-GODEFERD,mori01}. Very high rotation rate leads to an almost complete inhibition of nonlinear kinetic energy cascade, so that the anisotropy is reduced.

The growth of $\theta_b$ in time from the isotropic value is faster than the growth of $\theta_u$ for all finite values of $\Lambda$, indicating a faster structuring of the magnetic field towards a quasi-twodimensional state. A possible explanation lies in the nature of the phenomenon responsible for the growth of magnetic energy. Given that Alfv\'en waves are linear, the anisotropic growth of the magnetic energy is much faster than the slow nonlinear anisotropic energy transfer due to the Coriolis force.
Note that the two-dimensionalization of both the velocity and the magnetic fields is only partial since the Shebalin angles are far from the extremal value of $90^{\textrm{o}}$, characterizing a field invariant in the vertical direction.

\subsubsection{Angle dependent spectra of two-point correlations}

The Shebalin angle is a simple scalar characterizing the global anisotropy of the flow because of the integration over all wave vectors in equation \eqref{eq:sheb}.
In order to describe the anisotropy over all scales of the flow, it is necessary to introduce a scale- and angular-dependent quantity.
The angular energy spectrum is a very useful quantity in the axisymmetric case.
Contrary to classical energy spectra already presented in figures \ref{fig:comp_spec_1}, the energy density spectrum $e(\bm{k})$ is not spherically averaged, leading to $E(k)=\int e(\bm{k})\mathrm{d}^3\bm{k}$, but the angular dependence on the polar angle $\theta$ is conserved :
\begin{equation}
\label{eq:ekt_def2}
E_K(k,\theta)=\left[\int_{\theta-\Delta\theta/2}^{\theta+\Delta\theta/2}\cos\theta\mathrm{d}\theta\right]^{-1}\sum_{\substack{k-\Delta k/2<|\bm{k}|<k+\Delta k/2 \\ \theta-\Delta\theta/2<\theta<\theta+\Delta\theta/2}} \hat{u}_i(k,\theta)\hat{u}^*_i(k,\theta) \ .
\end{equation}
where $\Delta k$ and $\Delta\theta$ specify the discretization steps in Fourier space used for computing the anisotropic spectra (see figure \ref{fig:spang_lukas}, the blue region corresponds to the scales which contribute to $E(k,\theta)$).
Here, $\Delta k=1$ and $\Delta\theta=\pi/10$, these figures are necessarily linked to the DNS resolution, to ensure good statistical sampling.
In the isotropic case, all angular spectra collapse.

In addition to angular dependence, it is convenient to introduce the Craya-Herring frame ($\bm{e}^{(1)},\bm{e}^{(2)},\bm{e}^{(3)}$) \citep{saga08} to geometrically decompose the velocity field.
Because of the incompressibility condition ${\boldsymbol{\nabla}}{\boldsymbol{\cdot}}\bm{u}=0$ and because of ${\boldsymbol{\nabla}}{\boldsymbol{\cdot}}\bm{b}=0$, the velocity and the fluctuating magnetic field spectral components along $\bm{e}^{(3)}$ are identically zero since $\bm{k}{\boldsymbol{\cdot}}\hat{\bm{u}}=0$ and $\bm{k}{\boldsymbol{\cdot}}\hat{\bm{b}}=0$. Any spectral component can therefore be decomposed as a toroidal contribution (along $\bm{e}^{(1)}$) and a poloidal contribution (along $\bm{e}^{(2)}$). This representation is linked to the decomposition between shear (poloidal) and pseudo (toroidal) Alfv\'en waves \citep{bigo08}.

\begin{figure}
\unitlength 0.6mm
\begin{picture}(250,190)
        \put(40,100){\includegraphics[height=95\unitlength]{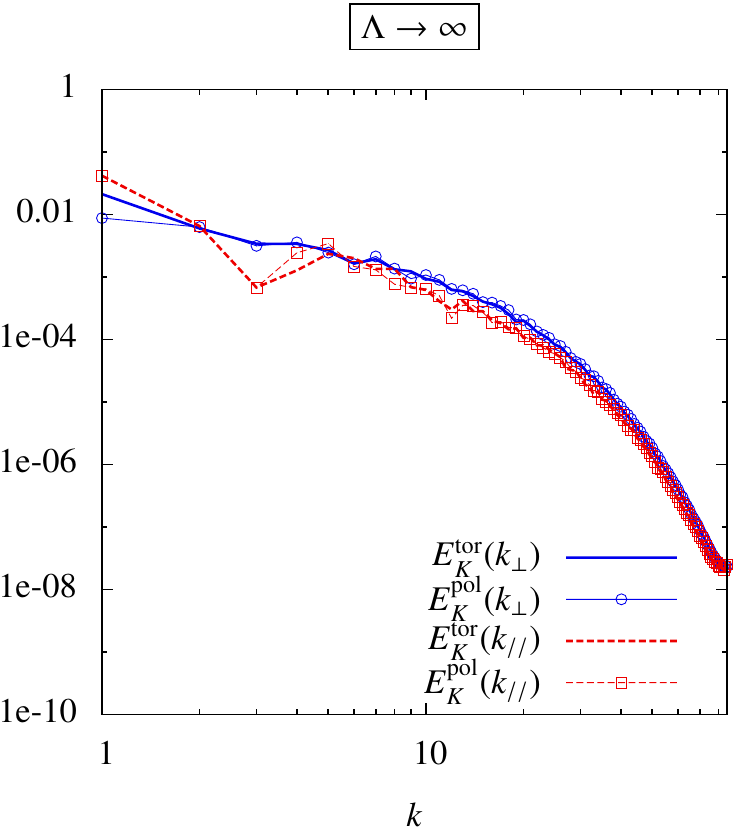}}
        \put(140,100){\includegraphics[height=95\unitlength]{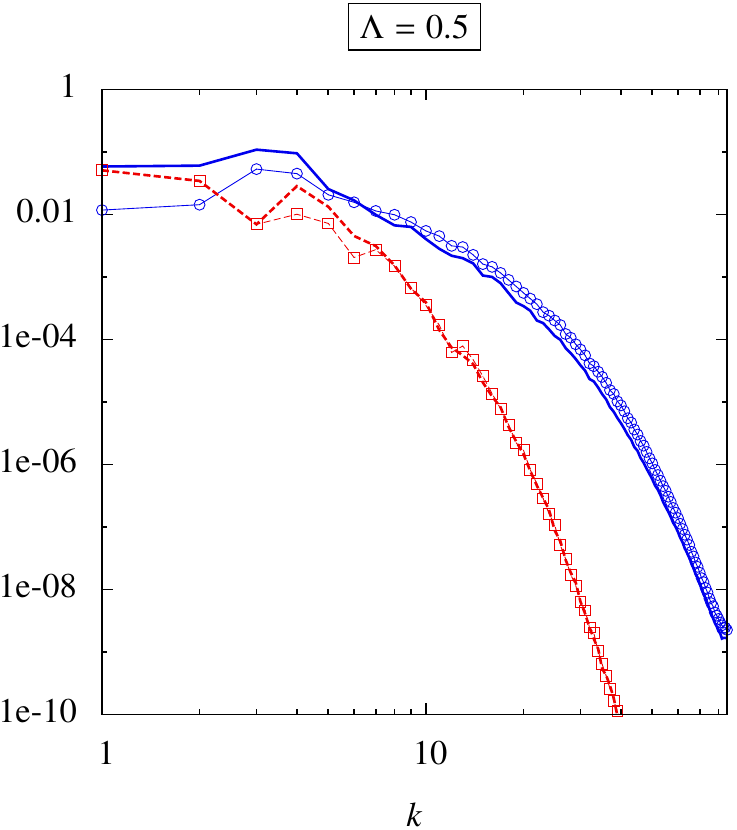}}
        \put(40,0){\includegraphics[height=95\unitlength]{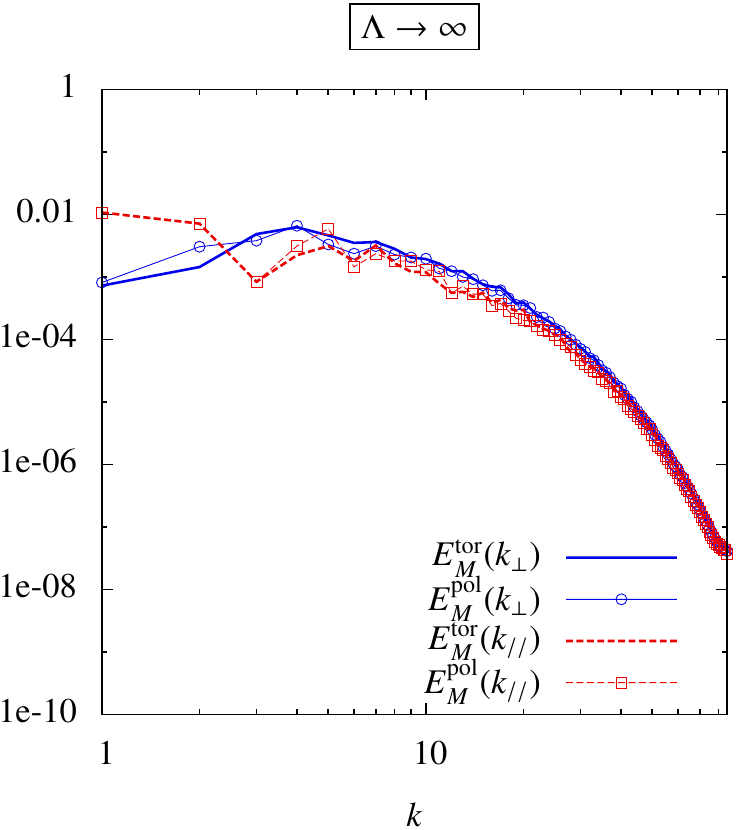}}
        \put(140,0){\includegraphics[height=95\unitlength]{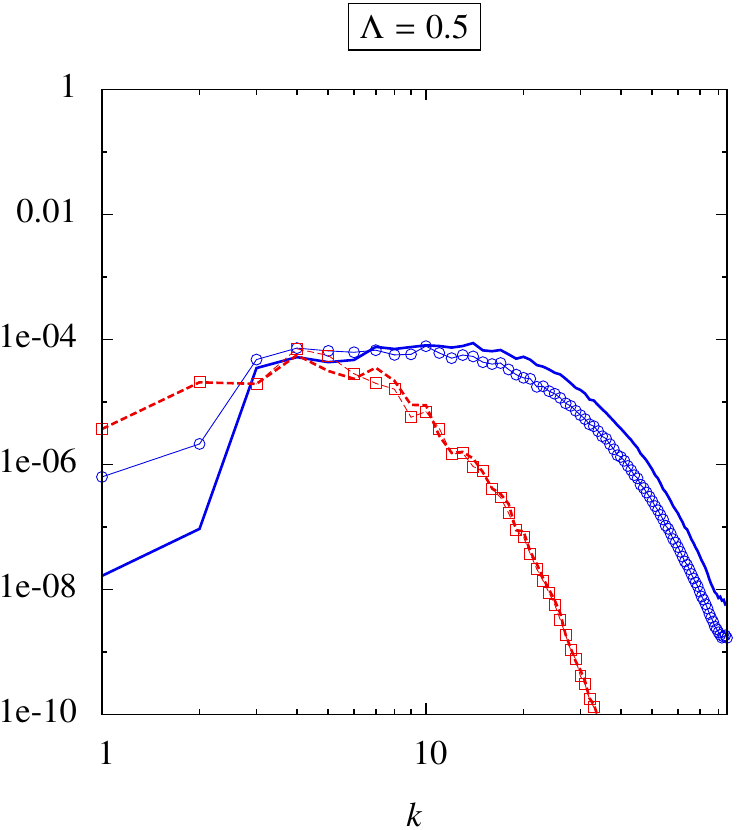}}
	\put(112,178){(a)}
	\put(210,178){(b)}
	\put(108,77){(c)}
	\put(206,77){(d)}
\end{picture}
\caption[]{(Color online) Angular energy spectra at $t^*\approx5$. Red color (or dotted lines) corresponds to the pole (\textit{i.e.} modes such that $\theta\approx0$) whereas blue color (or continuous lines) corresponds to the equator (\textit{i.e.} modes such that $\theta\approx\pi/2$). The lines without symbols correspond to the toroidal energy whereas the lines with symbols represent the poloidal energy. (a) Angular kinetic energy spectra without rotation. (b) Angular kinetic energy spectra with rotation at the same time. (c) Angular magnetic energy spectra without rotation. (d) Angular magnetic energy spectra with rotation at the same time.}
\label{fig:ekt_rot_mhd_hrm}
\end{figure}
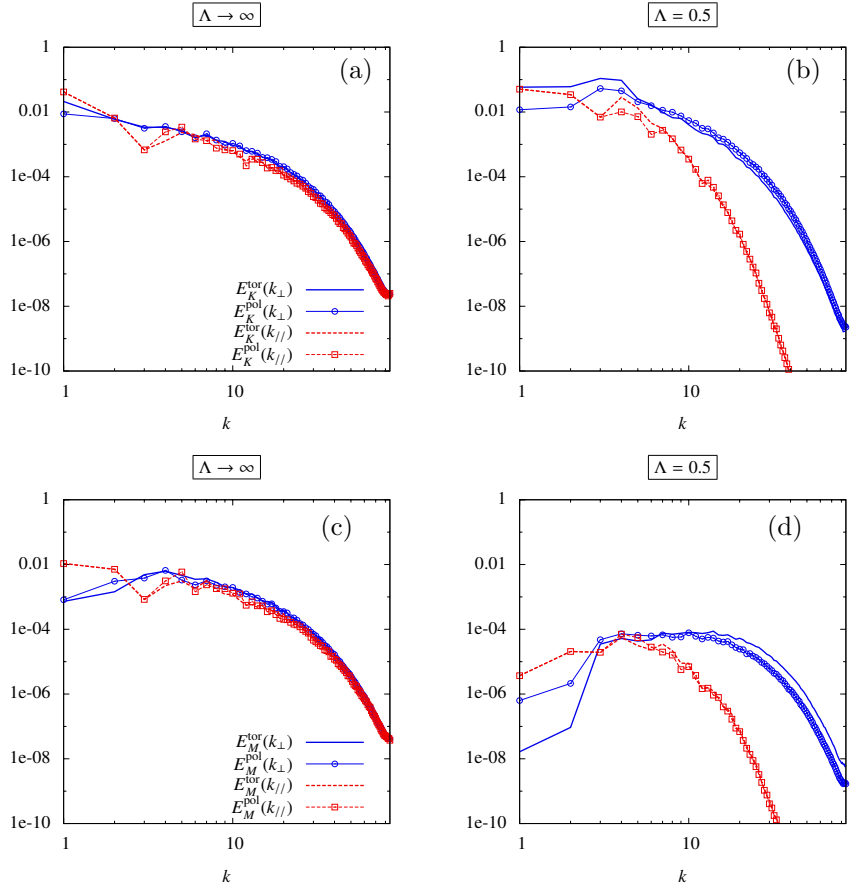

The angular energy spectra are plotted in figure \ref{fig:ekt_rot_mhd_hrm} at $t^*\approx5$ with and without rotation, for $\Lambda\rightarrow\infty$ and $\Lambda=0.5$.
The equator (modes such that $\bm{k}\perp\bm{\Omega}$) corresponds to blue solid lines whereas the pole (modes such that $\bm{k}\parallel\bm{\Omega}$) corresponds to red dotted lines.
For the sake of clarity, only polar and equatorial modes are represented (the spectra at intermediate angles $0<\theta<\pi/2$ lie monotonously between the spectra at $\theta=0$ and $\theta=\pi/2$). 
The lines without symbols correspond to toroidal spectra whereas the lines with symbols correspond to poloidal spectra.
In the following, we focus on the inertial and small scale dynamics (\textit{i.e.} $k>6$) since the statistical sampling for computing the angular-dependent spectra at low wave numbers is too coarse.

Let us first consider the velocity spectra.
In absence of rotation, figure \ref{fig:ekt_rot_mhd_hrm}(a) shows that the velocity field is roughly isotropic at all scales since all directional spectra almost collapse: there is equipartition of kinetic energy between polar and equatorial modes and equipartition of energy between poloidal and toroidal components.
One observes however a small increase of equatorial energy with respect to the polar one in the inertial range, which is consistent with a Shebalin angle slightly greater than its isotropic value, already seen on figure \ref{fig:sheb_hrm}(a).

With rotation, figure \ref{fig:ekt_rot_mhd_hrm}(b) shows that the anisotropy of the velocity field is similar to the one observed in non-magnetized rotating flows.
Without discriminating poloidal and toroidal contributions, one first observes a concentration of kinetic energy at the equator.
Surprisingly, this anisotropy increases with the wave number, which is apparently inconsistent with a possible restoration of isotropy at small-scales, even if one accounts for the low Reynolds number of the simulations.
By analogy with the Ozmidov scale in stratified fluids, some authors \citep{zema94,zhou95} introduce a cut-off wave number
\begin{equation}
k_{\Omega}=\sqrt{\frac{\Omega^3}{\epsilon_K}}
\end{equation}
where $\epsilon_K$ is the kinetic dissipation rate.
When $k>k_{\Omega}$, the effect of the Coriolis force may be neglected and a restoration of isotropy is expected.
When $k<k_{\Omega}$, anisotropic effects are dominant.
In figure \ref{fig:ekt_rot_mhd_hrm}(b), $k_{\Omega}\approx280$ so that, based on this approach, all scales are dominated by rotation.
Thus, higher Reynolds number and numerical resolution would be required to observe the restoration of isotropy at small scales.

Let us now distinguish the toroidal kinetic energy spectrum (solid lines) from the poloidal kinetic energy spectrum (lines with symbols).
This decomposition is of particular interest for the equatorial modes.
At the pole, axisymmetry imposes the poloidal and toroidal components to be equal.
One indeed observes equipartition between poloidal and toroidal energy for polar modes but not for equatorial modes, for which the toroidal energy is dominant at large scales whereas the poloidal energy is dominant at small scales.
This kind of anisotropy has already been observed in MHD turbulence at very low magnetic Reynolds number using the quasi-static approximation \citep{favier10}.
In that case, anisotropic Joule dissipation leads to a flow invariant in the direction of the imposed magnetic field.
From that quasi-twodimensional state in which $\partial/\partial z\approx0$ but $u_z$ is non necessarily zero, the non-linear cascade of energy can be very different depending on the component considered.
The vertical component of velocity behaves like a passive scalar (with a classical cascade) whereas the horizontal component behaves like purely 2D turbulence and thus displays a weak direct cascade and a possible inverse cascade in the absence of forcing.
This observation is consistent with the dominance of toroidal energy at small scales, with a horizontal toroidal mode, as shown on figure \ref{fig:spang_lukas}.
However, in rotating turbulence, the mechanisms responsible for the quasi-twodimensionalization differ from linear dissipative effects of quasi-static MHD turbulence.
The transition from 3D to quasi-2D is triggered by non-linear energy transfers and is thus less effective \citep{camb01}.
The previous scenario for quasi-static MHD turbulence, valid in low $R_M$ turbulence, is nonetheless very promising in the present case.

Let us now discuss the magnetic energy spectra depicted on figures \ref{fig:ekt_rot_mhd_hrm}(c) and (d).
Without rotation, the equipartition of energy is observed and the anisotropy of the velocity and magnetic fields are very similar.
With rotation, one observes again the attenuation of the magnetic energy at large scales, and equipartition of energy at small scales.
Thus, the anisotropy of the small-scale velocity field imposes the anisotropy of the small-scale magnetic field, with dominant equatorial energy.
The fluctuating magnetic field is therefore nearly invariant in the vertical direction, which is also consistent with the previous observation concerning the increase of the Shebalin angle $\theta_b$.
Note however that, focusing on equatorial modes (blue lines), the fluctuating magnetic field is characterized by a dominant poloidal energy at large scales and a dominant toroidal energy at small scales, which is the opposite of the velocity field.

\begin{figure}
\unitlength 0.6mm
\begin{picture}(200,120)
        \put(63,0){\includegraphics[height=120\unitlength]{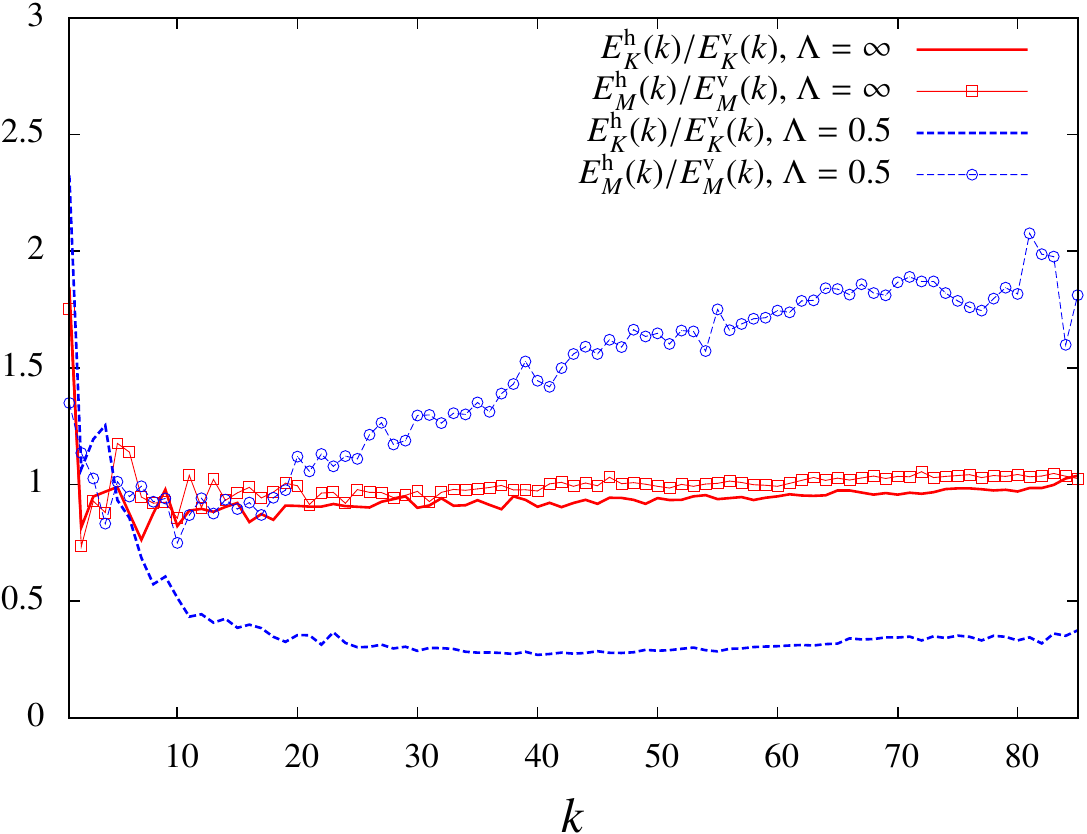}}
        \put(227,53){$\Lambda\rightarrow\infty$}
        \put(227,45){Non-rotating results}
        \put(221,49){\Big{\}}}
\end{picture}
\caption{(Color online) Scale-by-scale ratio between horizontal and vertical energies at time $t^*\approx5$. The lines represent the kinetic energy spectra whereas the lines with symbols represent magnetic energy spectra.
}
\label{fig:eii}
\end{figure}
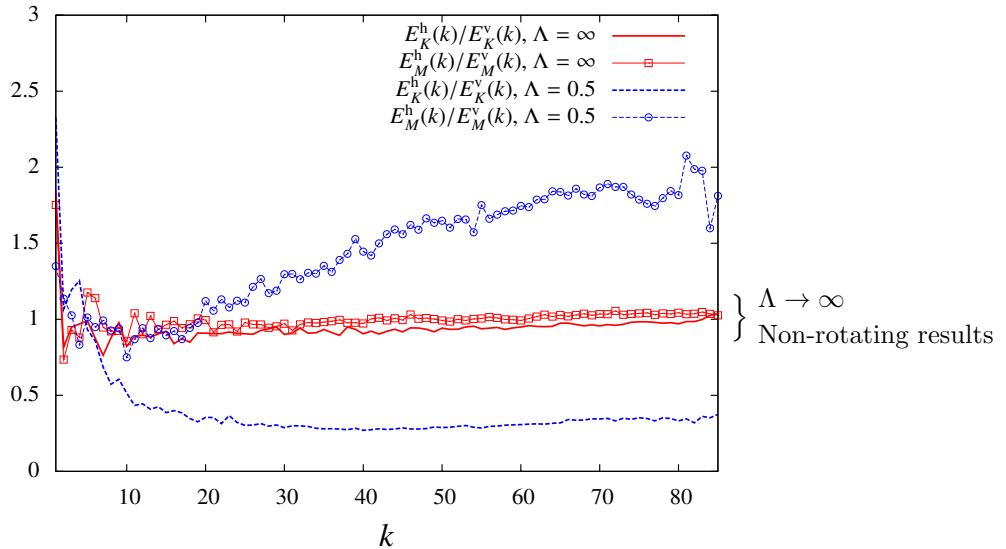

In the rotating cases, both velocity and magnetic energies are concentrated inside equatorial modes. However, for a given wavevector $\bm{k}\perp\bm{\Omega}$, the Fourier component can be vertical (\textit{i.e.} all the energy is therefore poloidal) or horizontal (\textit{i.e.} all the energy is toroidal, see figure \ref{fig:spang_lukas}), or in between (with mixed poloidal and toroidal contributions). The previous analysis seems to indicate that the small scale velocity field is dominated by its vertical component (the poloidal energy being slightly dominant on figure \ref{fig:ekt_rot_mhd_hrm}(b)) whereas the small scale magnetic fluctuations are dominated by their horizontal components (since the toroidal energy is dominant on figure \ref{fig:ekt_rot_mhd_hrm}(d)). This conclusion is confirmed when looking at the spherically averaged spectra, computed component by component. We separate the vertical energy spectrum
\begin{equation}
E_K^{\mathrm{v}}(k)=\sum_{k-\Delta k \le |\bm{k}| < k+\Delta k }\hat{u}_3(\bm{k})\hat{u}^*_3(\bm{k})
\end{equation}
from the horizontal energy spectrum
\begin{equation}
E_K^{\mathrm{h}}(k)=\frac{1}{2}\sum_{k-\Delta k \le |\bm{k}| < k+\Delta k }\big[\hat{u}_1(\bm{k})\hat{u}^*_1(\bm{k})+\hat{u}_2(\bm{k})\hat{u}^*_2(\bm{k})\big]
\end{equation}
where the index $1$ and $2$ refer to the directions perpendicular to $\bm{B}_0$ and $\bm{\Omega}$, and the index $3$ refers to the vertical direction. Similar definitions apply for the magnetic fluctuations.
Figure \ref{fig:eii} presents the scale-by-scale ratio between horizontal and vertical energy spectra of magnetic and velocity fluctuations. Without rotation (blue color), $E_K^{\mathrm{h}}(k)/E_K^{\mathrm{v}}(k)$ and $E_M^{\mathrm{h}}(k)/E_M^{\mathrm{v}}(k)$ are about unity at all scales. For $\Lambda=0.5$ (red color), one indeed observes that the vertical kinetic energy is dominant in the inertial range (\textit{i.e.} $E_K^{\mathrm{h}}/E_K^{\mathrm{v}}<1$) whereas the horizontal magnetic energy is dominant (\textit{i.e.} $E_M^{\mathrm{h}}/E_M^{\mathrm{v}}>1$). This confirms quantitatively our previous observations based on the poloidal-toroidal decomposition.

In summary, the anisotropy of the velocity field for rotating MHD turbulence at high magnetic Reynolds number is very similar to the anisotropy observed in non-magnetized rotating flows.
The non-linear angular transfers due to rotation accumulate the energy toward the equatorial plane. This quasi-twodimensionalization leads to a departure from the equipartition between toroidal and poloidal energies.
The fluctuating magnetic field is damped at large scales due to inertial waves, and is strongly anisotropic at small scales. The small scale velocity field is dominated by vertical motion, whereas the small scale magnetic field is mostly horizontal. 
%
%
\section{Conclusion}
We present in this paper data from direct numerical simulations of homogeneous incompressible turbulence submitted to both Coriolis and Lorentz forces.
This type of flows are of geophysical and astrophysical interests but are however characterized by numerous dimensionless parameters.
We therefore focus on the large magnetic Reynolds number, moderate interaction parameter and small Rossby number regime.

We have focused here on the study of the turbulent induction.
In that case, the equipartition between kinetic and magnetic energy due to Alfv\'en waves is broken by inertial waves introducing a separation scale, which depend on the Lehnert number.
This departure from equipartition leads to a reduction of the large scales magnetic energy in aid of the kinetic energy.
Concerning anisotropy, which is a key element in these flows, the velocity field is very similar to the well documented rotating hydrodynamic turbulence (because the anisotropy due to the imposed magnetic field is small), with a concentration of energy in the modes perpendicular to the rotation axis, which corresponds to the two-dimensional manifold in physical space.
The magnetic fluctuations display the same kind of angular anisotropy as the velocity field (\textit{i.e.} dominant equatorial energy at small scales), but with a different repartition between poloidal and toroidal components: the small-scale velocity field is dominated by poloidal modes (as in hydrodynamic rotating turbulence) whereas the small-scale magnetic field is dominated by toroidal modes.

Many of the features of rotating MHD turbulence can be explained thanks to the linear analysis. Some basic properties of magneto-inertial waves (damping of the magnetic energy, misalignment between $\bm{u}$ and $\bm{b}$) are of interest to study the fully nonlinear turbulent regime.

The small magnetic Reynolds is also of interest but the dynamics are in that case much more complex since both Coriolis force and Joule dissipation induce strong anisotropy on the flow. The angular transfer of energy due to Coriolis force compete with the anisotropic Joule dissipation, for example considering the quasi-static approximation.
These two effects tend to concentrate kinetic energy on equatorial modes (\textit{i.e.} modes such that $\bm{k}\perp\bm{\Omega}$) and interesting dynamics could arise depending on the Elsasser number.
The weak turbulence state (\textit{i.e.} $B_0\gg u_0$), considered in previous studies \citep{bigo08}, could also be of interest.

An other limitation of the present paper comes from the direction of the vector rotation $\bm{\Omega}$ with respect to the imposed magnetic field $\bm{B}_0$.
We focus here on the axisymmetric case and the perpendicular case (being the most plausible configuration inside the Earth's core) is postponed to future studies.
In that case, the competition between Alfv\'en and inertial waves is much more complex, depending not only on the scale and on the polar angle $\theta$ between $\bm{k}$ and $\bm{\Omega}$ (being vertical) but also on the azimuthal angle $\phi$ between $\bm{k}$ and $\bm{B}_0$ (being horizontal).

Another perspective concerns the forcing by physical instabilities. Decaying turbulence is chosen here not to alter the development of anisotropy. It is however possible to inject energy by mean of precessional instability (for instance by extending the theoretical study by \cite{sahli09} to DNS) or by unstable stratification using the homogeneous approach of \cite{ors97}.

Finally, initial conditions including initial magnetic fluctuations (with or without imposed magnetic field) is the next step to consider the effect of the Coriolis force on developed MHD turbulence. In that case, the simplified linear solutions \eqref{eq:NSrotspectralsol1} and \eqref{eq:NSrotspectralsol2} will depend on the initial condition for the magnetic mode, and a different dynamical regime is expected both from linear and nonlinear approaches.

The authors would like to acknowledge support of the IDRIS (Institut du D\'eveloppement et des Ressources en Informatique Scientifique) for computational time on NEC SX-8 under the project 082206.
\bibliographystyle{gGAF}
\bibliography{biblio}
\end{document}